\pdfoutput=1%
\documentclass[a4paper,american,floatfix,pdftex,superscriptaddress,twoside,citeautoscript
reprint,twocolumn,%,final% add final to see real layout, no todonotes anymore, ...
aps,pra%
]{revtex4-2}%

\PassOptionsToPackage{unicode}{hyperref} % please leave linkframes in the file for online viewing, arXiv, etc. -- they help the reader
\usepackage{amsfonts,amsmath,amssymb}
\usepackage{tabularx}
\usepackage{graphicx}
\usepackage[T1]{fontenc}
\usepackage[utf8]{inputenc}
\usepackage{hypernat}
\usepackage[ams,todos]{CMImacros}
\usepackage{verbatim}
\usepackage{multirow}
\usepackage[table]{xcolor}
\usepackage{colortbl}
\graphicspath{{./fig/}} %define figure directory

\newcommand{\cfeldesy}{\affiliation{Center for Free-Electron Laser Science CFEL, Deutsches
      Elektronen-Synchrotron DESY, Notkestraße 85, 22607 Hamburg, Germany}}%
\newcommand{\uhhchem}{\affiliation{Department of Chemistry, Universität Hamburg,
      Martin-Luther-King-Platz 6, 20146 Hamburg, Germany}}%
\newcommand{\uhhcui}{\affiliation{Center for Ultrafast Imaging, Universität Hamburg, Luruper
      Chaussee 149, 22761 Hamburg, Germany}}%
\newcommand{\uhhphys}{\affiliation{Department of Physics, Universität Hamburg, Luruper Chaussee 149,
      22761 Hamburg, Germany}}%
\newcommand{\stemail}{\email[]{sebastian.trippel@cfel.de}}%
\newcommand{\cmiweb}{\homepage{https://www.controlled-molecule-imaging.org}}%

\newcommand*{\CHHBrI}{\ensuremath{\text{CH}_2\text{BrI}}\xspace} % CH2BrI
\newcommand*{\CHHII}{\ensuremath{\mathrm{CH_2I_2}}\xspace} % CH2I2:?Oionz
\newcommand*{\CFFBrBr}{\ensuremath{\mathrm{CF_2Br_2}}\xspace} % CF2Br2

\newcommand*{\CHHIion}{\ensuremath{\mathrm{CH_2I^+}}\xspace} % CH2Iion

\newcommand*{\CFFII}{\ensuremath{\mathrm{CF_2I_2}}\xspace} % CF2I2
\newcommand*{\CFFIIast}{\ensuremath{\mathrm{CF_2I_2^{\ast}}}\xspace} %CF2I2 star
\newcommand*{\II}{\ensuremath{\mathrm{I_2}}\xspace}  %I2
\newcommand*{\IIstar}{\ensuremath{\mathrm{I_2^\ast}}\xspace}  %rotationally excited I2
\newcommand*{\CFFI}{\ensuremath{\mathrm{CF_2I}}\xspace}  %CF2I
\newcommand*{\CFFIstar}{\ensuremath{\mathrm{CF_2I^\ast}}\xspace}  %CF2I
\newcommand*{\Iatom}{\ensuremath{\mathrm{I}}\xspace}  %I atom
\newcommand*{\CFF}{\ensuremath{\mathrm{CF_2}}\xspace} %CF2

\newcommand*{\CFFIIion}{\ensuremath{\mathrm{CF_2I_2^+}}\xspace} %CF2I2 ion
\newcommand*{\IIion}{\ensuremath{\mathrm{I_2^+}}\xspace}  %I2 ion
\newcommand*{\IIstarion}{\ensuremath{\mathrm{I_2^{\ast +}}}\xspace}  %rotationally excited I2 ion
\newcommand*{\CFFIion}{\ensuremath{\mathrm{CF_2I^+}}\xspace}  %CF2I ion
\newcommand*{\Iion}{\ensuremath{\mathrm{I^+}}\xspace}  %I ion
\newcommand*{\CFFion}{\ensuremath{\mathrm{CF_2^+}}\xspace} %CF2 ion
\newcommand*{\CFion}{\ensuremath{\mathrm{CF^+}}\xspace} %CF ion

\newcommand*{\CFFIIwaterion}{\ensuremath{\mathrm{CF_2I_2 \cdot H_2O^+}}\xspace} %cf2i2.water ion
\newcommand*{\CFFIICFFIion}{\ensuremath{\mathrm{CF_2I_2\cdot CF_2I^+}}\xspace}%CF2I2.CF2I ion
\newcommand*{\CFFIwaterion}{\ensuremath{\mathrm{CF_2I\cdot H_2O^+}}\xspace}%CF2I.water ion
\newcommand*{\CFFIIIion}{\ensuremath{\mathrm{CF_2I_2\cdot I^+}}\xspace}%CF2I2.CF2I ion
\newcommand*{\oneBone}{\ensuremath{^1B_1}\xspace}%1B1 symmetry state
\newcommand*{\bCI}{\ensuremath{\mathrm{C\text{--}I}}\xspace} % C-I bond

\newcommand*{\ueV}{\ensuremath{\text{{\textmu{eV}}}}\xspace} % micro eV
\newcommand{\range}[2]{#1\nobreakdash--#2}

\begin{document}
\title{Rotational-state-controlled dissociative ionization dynamics in \CFFII}%
\author{Nidin Vadassery}\cfeldesy\uhhchem%
\author{Ivo S. Vinkl\'arek}\cfeldesy%
\author{Sebastian Trippel}\stemail\cmiweb\cfeldesy\uhhphys\uhhcui%
\author{Jochen Küpper}\cfeldesy\uhhchem\uhhphys\uhhcui%%%
\begin{abstract}\noindent
   We investigated
   strong-field dissociative ionization of \CFFII ensembles prepared in different initial
   rotational-state distributions using an electrostatic deflector. Pronounced changes in
   ion-channel branching ratios revealed a strong dependence of the fragmentation dynamics on the
   initial rotational excitation. Analysis of fragment yields and their laser-power dependences
   identifies resonance-enhanced multiphoton ionization through an intermediate excited state,
   \CFFIIast. The measured branching behavior indicates competition between stabilization into bound
   ionic states and dissociative channels, driven by near-threshold non-adiabatic Coriolis-type
   coupling. Tuning the rotational energy by only a few~\ueV is sufficient to significantly alter
   the ionization dynamics and to redistribute the reaction products. These findings demonstrate the
   key role of rotational excitation in controlling non-adiabatic dynamics following strong-field
   ionization of \CFFII.
\end{abstract}
\maketitle

\section{Introduction}
\label{sec:introduction}
Polyhalomethanes are well-recognized photolytic sources of reactive halogen species, playing a role
in aerosol chemistry and contributing to ozone depletion~\cite{Koenig:sciadv7:eabj6544,
ODowd:Nature417:632, Saiz-Lopez:CR112:1773}. An accurate description of such processes requires a
molecular-level understanding of the underlying photodissociation
dynamics~\cite{Gravestock:CPC11:3928}. In particular, reaction branching, product energy
partitioning, and nonadiabatic dynamics can depend sensitively on the initial molecular state, yet
these state-resolved effects remain largely unexplored~\cite{Suchan:FD212:307}.

Among polyhalomethanes, difluorodiiodomethane (\CFFII) is a particularly attractive and
experimentally accessible model system for probing state-dependent fragmentation
dynamics~\cite{El-Khoury:JPCA113:10767}. \CFFII serves as a prototype molecule for understanding
related atmospherically relevant halomethanes such as \CFFBrBr
(Halon-1202)~\cite{Liu:applsci11:1704}, \CHHII~\cite{Horton:JCP150:174201}, and
\CHHBrI~\cite{Recio:JPCA126:8404}, which exhibit comparable absorption spectra. However, despite the
structural and spectroscopic similarities between \CHHII and \CFFII~\cite{Wannenmacher:JCP95:986,
   Toulson:PCCP18:11091}, their \bCI bond dissociation energies differ substantially, 2.74~eV for
\CFFII compared with 4.94~eV for \CHHII. As a result, \CFFII exhibits multiple dissociation channels
at relatively low excitation energies, making it significantly less
photostable~\cite{Baum:JCP98:1999}.

Upon UV irradiation, neutral \CFFII exhibits electronic excitation to several dissociation channels,
as evidenced by state-selected photodissociation dynamics observed across
\range{248}{351}~nm~\cite{Baum:JCP98:1999}. At low excitation energies in the \range{337}{351}~nm
range, two-body dissociation dominates, primarily yielding iodine atoms in the ground spin--orbit
state $^2P_{3/2}$~\cite{Wannenmacher:JCP95:986}. With increasing excitation energy, three-body
fragmentation becomes more important, accompanied by enhanced production of spin--orbit excited
iodine atoms in the $^2P_{1/2}$ state~\cite{Bergmann:JCP109:474}. Beyond atomic-fragment channels,
time-resolved VMI studies of \CFFII with two-photon excitation at 264~nm revealed the delayed
elimination of highly rotationally excited \II~\cite{Radloff:CPL291:173, Farmanara:JCP113:1705,
   Roeterdink:JCP117:6500}, a product-state signature associated with roaming
dynamics~\cite{Townsend:Science306:1158}. Similar roaming-like dynamics and transient
$iso$-intermediates were observed for analogous iodine-containing halomethanes such as \CHHII and
\CHHBrI rather than prompt bond fission~\cite{Borin:PCCP18:28883, Recio:JPCA126:8404,
   Anderson:JCP139:194307, El-Khoury:JPCA113:10767}. In \CFFII, femtosecond spectroscopy revealed
the formation of $iso$-difluorodiiodomethane within $400$~fs following 350~nm
excitation~\cite{El-Khoury:JCP132:124501} suggestive of roaming dynamics.

A similar level of complexity has been observed in ionization-induced dynamics. Here, ultrafast
dissociation was observed on a sub-100~fs timescale through multiple channels, yielding atomic and
molecular iodine ions~\cite{Radloff:CPL291:173, Farmanara:JCP113:1705}. Subsequent time-resolved
mass-spectrometry measurement identified prompt \IIion formation at pump--probe delays $\leq250$~fs
and a delayed \IIion channel $\geq500$~fs, which was attributed to asynchronous molecular-iodine
elimination~\cite{Roeterdink:JCP117:6500}. Strong-field-ionization of \CHHBrI and \CHHII further
showed that fragmentation branching in halomethanes can be shaped by dynamic ionic resonances and
corresponding steering of nuclear motion in the molecular cation~\cite{Pearson:JCP127:131101}.
Together, these findings raise the question of the sensitivity of dissociation pathways to the
initially prepared state and to laser-interaction conditions.

Conventionally, it is assumed that product-state distribution is determined by the excitation energy
and the participating potential-energy surfaces~\cite{Crim:ARPC1:397}. An alternative route to
control the dissociation dynamics is through initial-state preparation, such as selecting
well-defined rotational quantum states prior to photoexcitation~\cite{Prlj:JPCA127:7400,
   Suchan:FD212:307}. This sensitivity to the initial-state distribution is supported by earlier
experimental and theoretical studies, which showed that reaction outcomes depend strongly on the
initial phase-space conditions and their mapping onto excited and ionic
states~\cite{Suchan:FD212:307, Bamford:JCP82:3032, Xie:JPCA104:1009}. In particular, the initial
nuclear phase space can play a decisive role in non-adiabatic dynamics and in the resulting
fragment-state distributions~\cite{Janos:ACR58:261, Vinklarek:PCCP23:14340}.

Non-adiabatic couplings play a central role in the photodissociation dynamics of halomethanes, where
electronic-state crossings can strongly influence internal energy
partitioning~\cite{Toulson:PCCP18:11091}. Velocity map imaging studies showed that carbon–halogen
bond cleavage can impart substantial torque to the fragments, leading to pronounced rotational
excitation~\cite{MurilloSanchez:PCCP20:20766, Recio:JPCA126:8404, Vinklarek:PCCP23:14340}. The
delayed cleavage of the second C–I bond observed in highly ionized \CHHII suggests that rotational
motion within the intermediate fragment dynamically stabilizes the transient potential well,
highlighting Coriolis coupling as an important mechanism connecting rotational dynamics with
dissociation pathways~\cite{Trost:jpbamop58:085101}. Consistent with this picture, recent studies
have shown that photodissociation branching ratios are highly sensitive to the initial rotational
state, particularly near the dissociation threshold, owing to enhanced non-adiabatic coupling among
spin–orbit–coupled dissociative states~\cite{Keyu:ultrafastsci4:0073, Jiang:JPC156:191101}. Similar
effects may also govern non-adiabatic transitions along dissociative coordinates in related
systems~\cite{Liu:PRL136:053201}, motivating state-selected experiments to directly probe the role
of rotational energy in product branching and dissociation dynamics.

Controlled rotational-state ensembles can be prepared experimentally by manipulating molecular beams
with electric fields~\cite{Meerakker:CR112:4828, Chang:IRPC34:557}. Electrostatic deflection
spatially disperses polar molecules according to their Stark interaction, thereby enabling state-,
size-, and isomer-selected molecular samples~\cite{Chang:CPC185:339,
   Trippel:RSI89:096110,Filsinger:PRL100:133003, He:RSI95:113301}. The spatially dispersed molecular
beam retains distinct rotational-state distributions across its transverse vertical profile,
providing well-defined initial conditions for subsequent photoexcitation and reaction-dynamics
studies~\cite{Chang:Science342:98, Vinklarek:JPCA128:1593, Johny:PCCP26:13118, Kilaj:NatComm12:6047,
   Kilaj:NatComm9:2096}.

Here, we present our observations from single-pulse strong-field ionization of rotational-state
dispersed \CFFII. We present the resulting mass spectra, highlighting the dominant ionic fragments
produced during strong-field ionization. We then analyze the spatial separation achieved \emph{via}
electrostatic deflection and extract the corresponding rotational state distributions. Next, we
examine the variation of fragment branching ratios across the spatially deflected molecular beam,
together with laser-power-dependent measurements that reveal the effective ionization order of the
observed fragments as a function of deflection. Finally, we discuss possible mechanisms, including
non-adiabatic effects, underlying the observed changes in reaction pathways induced by the initial
rotational-state distribution and present a heuristic model describing the branching ratios as a
function of rotational energy in \CFFII.

\section{Methods}
\label{sec:methods}
For the measurements, we used the recently described endstation for controlled molecule experiments
eCOMO~\cite{Jin:RSI96:023305}. \CFFII (Apollo Chemical, UK) was co-expanded with helium
($p_{\mathrm{He}} = 100$~bar) into the source vacuum chamber using a pulsed Even–Lavie
valve~\cite{Even:EPJTI2:17} operated at 295~K and a repetition rate of 100~Hz. The expanded gas was
introduced into the deflector chamber through a 3~mm skimmer as a molecular beam. Subsequently, the
molecular beam was skimmed again using a 1.5~mm skimmer positioned in front of a $b$-type
electrostatic deflector operated at a voltage of 24~kV. Exploiting the Stark effect, the molecules
experienced a force in the inhomogeneous electric field, resulting in a spatial deflection
proportional to their effective-dipole-moment-to-mass ratio~\cite{Chang:IRPC34:557}. Next, a
knife-edge was employed to enhance the \CFFII column density in the interaction
region~\cite{Trippel:RSI89:096110}. Finally, the deflected molecular beam was transported through an
1.5~mm skimmer into the experimental chamber equipped with an ion spectrometer. The molecular beam
was intersected by a Ti:sapphire laser (Astrella, Coherent) with a central wavelength of 800~nm,
focused to a 54.75~\um diameter FWHMI spot, and with a pulse duration of 45~fs. The laser pulses
induced strong-field ionization of \CFFII, which has a vertical ionization energy of
$\Ei=9.8$~eV~\cite{Radloff:CPL291:173, Farmanara:JCP113:1705}.

The spectrometer was used in two modes: velocity-map imaging (VMI) and spatial-map imaging (SMI).
VMI was employed to extract the velocity distributions of the fragment ions, whereas SMI was
utilized to determine ion yields. The use of SMI mitigated detector saturation, which otherwise
occurred when multiple ions impinged on a localized area of the detector simultaneously. The
detector consisted of a microchannel plate $z$-stack and a phosphor screen. Each ion impact on the
detector yielded a short flash of light recorded by a Timepix3
camera~\cite{Bromberger:JPB55:144001}, providing position-of-arrival and time-of-flight for
individual events.

The data were recorded at various laser intensities in the range \range{20}{50}~\TWpcmcm to
investigate the effective ionization order $n$ for \CFFII ion channels~\cite{LHuillier:PRA27:2503,
  Wiese:NJP21:083011}. The Keldysh parameter for the given peak laser intensities changes from 2.13
at 20~\TWpcmcm to 1.35 at 50~\TWpcmcm, indicating that multiphoton ionization is the dominant
process.  The effective ionization orders, $n$, associated with the measured ion yields were
calculated from linear fits to the log–log plots of ion yield as a function of peak laser intensity
using the power-law relationship $Y_{\mathrm{ion}}(I) \propto I^n$, where $Y_{\mathrm{ion}}$ denotes
the ion yield at a given peak laser intensity $I$.

\section{Results}
\label{sec:results}
\subsection{\CFFII dissociative ionization}
\label{sec:mass_spectrum}
\begin{figure*}
  \includegraphics[width=\textwidth]{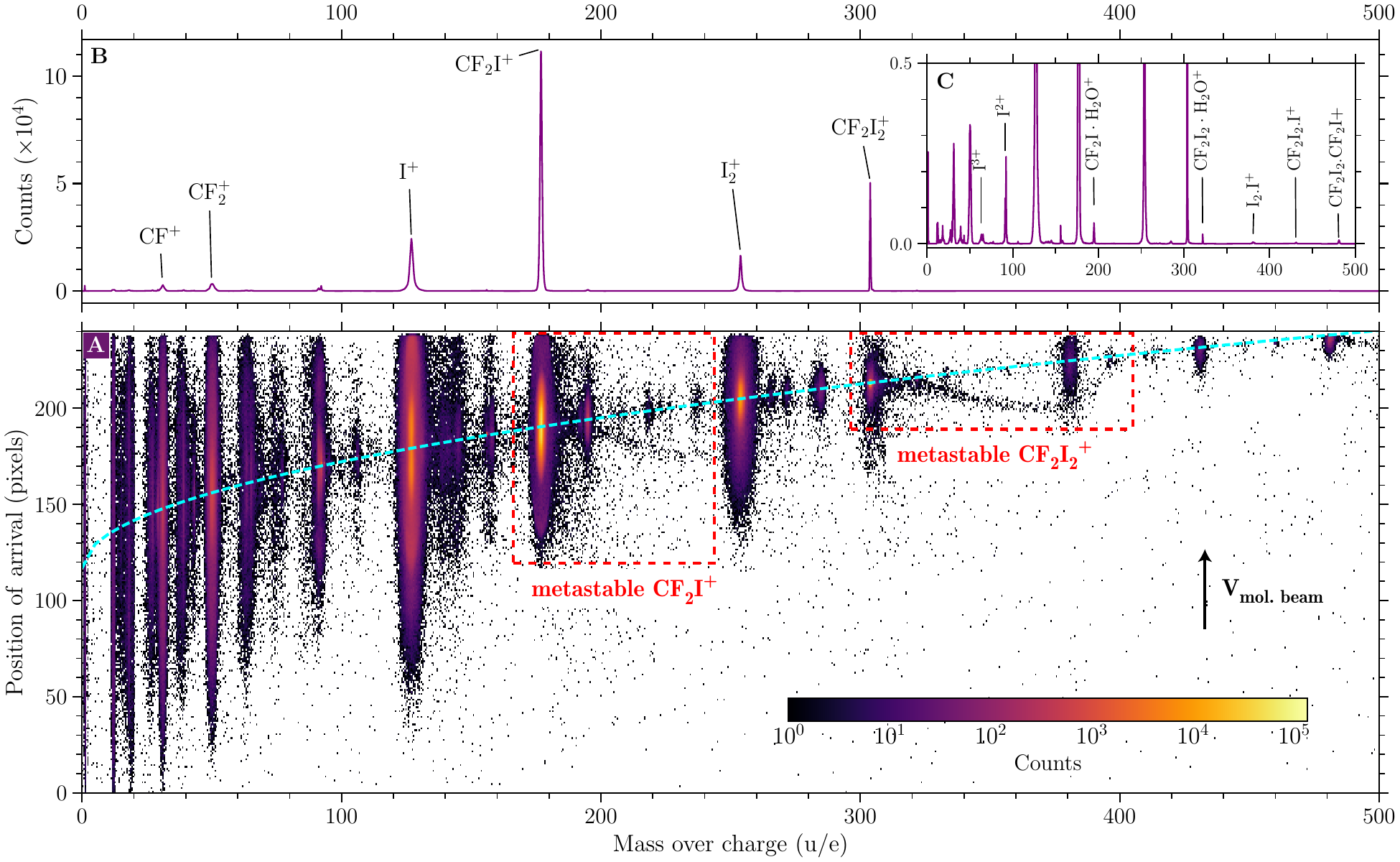}
  \caption{\textbf{A}: Background-subtracted correlation map of fragment ions recorded in VMI mode
    as a function of mass-to-charge ratio and arrival position on the detector. The dashed cyan line
    traces the vertically displaced image center arising from the molecular-beam velocity.
    \textbf{B}: Mass spectrum of the fragmentation products obtained from the integration of all
    signal in panel A. \textbf{C}(inset): Vertically magnified view emphasizing low-intensity
    contributions. The red box in panel A highlights weak metastable fragmentation
    channels~\cite{Vinklarek:SlowFragmentationDynamics:inprep}, visible as signal tails associated with \CFFIIion
    and \CFFIion.}
  \label{fig:cf2i2_mass_spec}
\end{figure*}

\begin{figure}
  \centering
  \includegraphics[width=\linewidth]{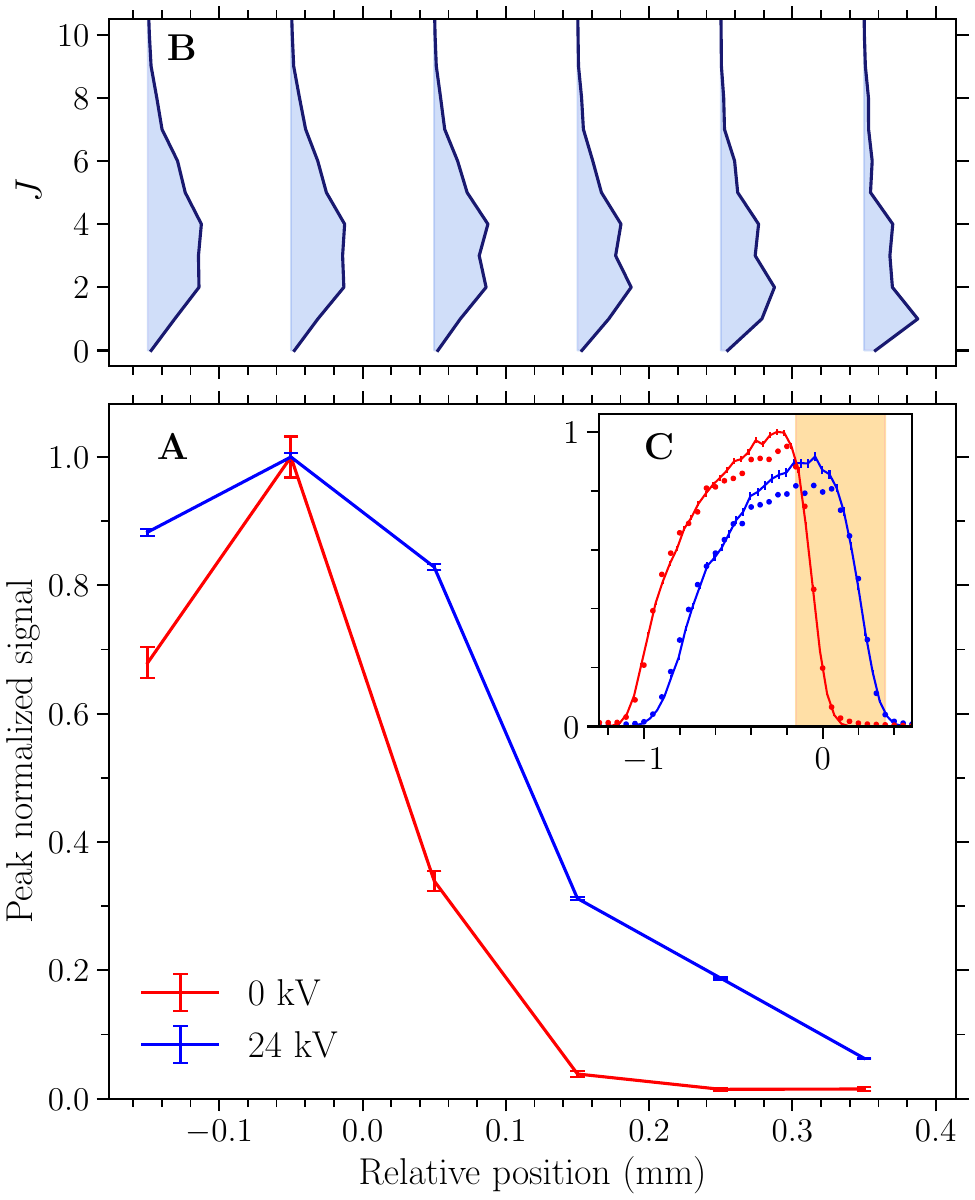}%
  \caption{\textbf{A}: Spatially resolved profiles of the deflected/dispersed (blue) and
     undeflected/direct (red) molecular beams measured within the orange shaded region of the total
     molecular-beam profile shown in panel C. \textbf{B}: Rotational-state distributions
     corresponding to the relative positions within the deflected beam shown in panel A.}
  \label{fig:deflection-profile}
\end{figure}

An overview of all acquired ions from strong-field ionization of \CFFII, recorded in VMI mode, is
presented in the lower panel of \autoref{fig:cf2i2_mass_spec}.
The data were recorded at a peak intensity of 31.4~\TWpcmcm and a relative position of 0.35~mm in
the deflected molecular beam (\autoref{fig:deflection-profile}). The 2D representation maps the
background-subtracted ion yield against the mass-to-charge ratio $m/q$ and arrival position along
the molecular beam propagation direction. The $m/q$ values were obtained from a quadratic
transformation of time-of-flight ($m/q\propto t^{2}$). The signal associated with each $m/q$
contains information equivalent to an individual velocity-map image, whose center is vertically
displaced due to the molecular-beam velocity. For VMIs of individual fragments, see Supplementary
Information. The ion signal out of the dashed cyan line indicates kinetic energy gained during the
fragmentation process. The upper panel shows the mass spectrum, obtained by integrating along the
position-of-arrival axis, with an inset displaying low-yield fragmentation channels.

The major ion channels observed following strong-field ionization of \CFFII are the parent ion
\CFFIIion ($\mq{304}$), \IIion ($\mq{254}$), \CFFIion ($\mq{177}$), \Iion ($\mq{127}$), and \CFFion
($\mq{50}$). This is illustrated in the ion image by broad features corresponding to the statistical
distributions of ion kinetic energies. We also observe minor contributions from clusters, including
\CFFII homomers and heteromers formed with water above \mq{304}, as illustrated in the mass
spectrum. The ion channels originating from heteromers are \CFFIIwaterion (\mq{322}) and
\CFFIwaterion (\mq{195}). Ion channels attributed to \CFFII homomers include \CFFIICFFIion
(\mq{481}) and \CFFIIIion (\mq{431}). Clusters with higher mass ($m/q>500~\mathrm{u/e}$) were not
detected because they fell outside of the detector area. Based on the relative intensities, we
estimate the signal contributions from all clusters to $\smaller1\%$. Due to the kinetic energy
distributions of the fragments, some ions fail to hit the detector's sensitive area, even for low
mass. Consequently, the ion yields are slightly underestimated in these cases. With an estimated
loss of $<0.05$\% the associated systematic uncertainty is negligible. In addition, we also observed
multi-charged ion contributions in the order of $0.1\%$, represented by $\mathrm{I^{2+}}$
(\mq{63.5}) and $\mathrm{I^{3+}}$ (\mq{42.3}). A minor contribution from metastable fragmentation
channels was recorded, as revealed by signal tails of \CFFII (\mq{304}) and \CFFIion ($\mq{177}$) in
the ion map \autoref{fig:cf2i2_mass_spec}~\cite{Vinklarek:SlowFragmentationDynamics:inprep}. Similar
long-lived \CFFIIion states were observed after multiphoton ionization at a laser wavelength of
264~nm~\cite{Roeterdink:JCP117:6500}. A precise assignment to specific rovibronic states of
\CFFIIion, however, lies beyond the scope of the present work.

\subsection{Electrostatic deflection of \CFFII}
\label{sec:deflection}
Electrostatic deflection was used to disperse the molecular beam and to prepare \CFFII ensembles
with varying rotational-state populations across the vertical molecular beam
profile~\cite{Filsinger:JCP131:064309, Trippel:RSI89:096110, Chang:IRPC34:557}. The
background-subtracted experimental and simulated deflection profiles for \CFFII are shown in
\autoref[C]{fig:deflection-profile}.
The red and blue lines represent profiles of the undeflected and deflected molecular beams,
respectively. The profiles were obtained from all detected ions and are peak-normalized with respect
to the undeflected profile. We observe a shift of the signal intensity by $\sim0.27$~mm at the
half-height (50\%) level of maximum signal in the presence of the inhomogeneous electric field. The
rotational temperature (\Trot) of the whole \CFFII ensemble was estimated to be 0.51(10)~K from
comparison with our molecular trajectory simulations~\cite{Chang:CPC185:339}. From the same
simulation, we obtained the rotational-state distribution and the mean rotational quantum number
$\langle{J}\rangle$ of \CFFII at different positions across the molecular beam as shown in
\autoref[B]{fig:deflection-profile}. The electrostatic deflection results in a decrease in
$\langle J\rangle$, corresponding to progressively colder rotational ensembles with increasing
relative position. For the present experimental conditions, $\langle{J}\rangle$ decreases from
3.94(3) to 2.70(14) as a function of increasing relative position and corresponding spatial
separation.

Due to adiabatic and non-adiabatic field rotations during molecular propagation, the $M$-state
distribution cannot be reliably evaluated in our current setup and is thus not considered in this
analysis. Specifically, the electric field in the deflector is oriented perpendicularly to the field
in the VMI, causing the quantization axis to rotate by 90$^\circ$. Consequently, any eigenstate
defined by a magnetic quantum number $M$ with respect to one axis would appear close to a fully
mixed superposition with respect to the other~\citep{Veldhoven:PRA66:032501}. We therefore assume an
uniform $M$-state distribution.

\subsection{Ionization order}
\label{subsec:power_dependene_cf2i2}
\begin{figure}
  \centering
  \includegraphics[width = 1.0\linewidth]{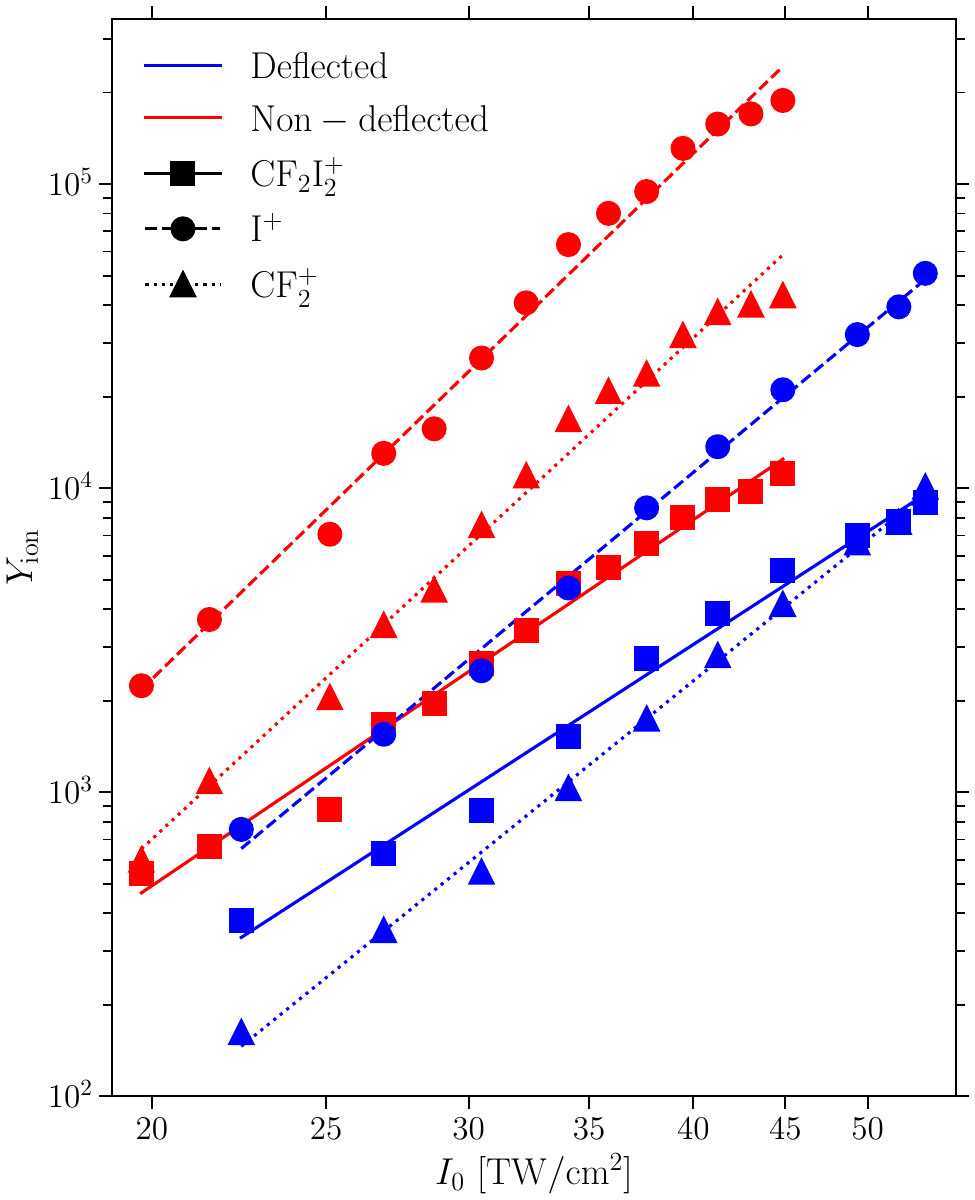}%
  \caption{Log-log plots of the ion yield dependence on the laser peak intensity for \CFFIIion,
     \Iion, and \CFFion, measured for the non-deflected ($-0.15$~mm, red) and deflected (0.35~mm,
     blue) positions of the molecular beam. The straight lines represent linear regression fits and
     the extracted slopes are tabulated in \autoref[columns 5, 6]{tab:reaction-scheme}.}
  \label{fig:log-plot-energy-level}
\end{figure}

The power dependence of the major ion channels was investigated at the non-deflected ($-0.15$~mm)
and deflected (0.35~mm) positions of the molecular beam for laser intensities of 20...50~\TWpcmcm.
The ion yield on a log–log scale exhibits a positive linear dependence on the laser peak intensity
for all ion channels. \autoref{fig:log-plot-energy-level} shows this behavior for selected ion
signals of \Iion, \CFFion, and \CFFIIion.
All power dependencies were fitted with straight lines to estimate the effective number of photons
required for the appearance of the individual ions. The fit results for the selected ions are shown
in \autoref{fig:log-plot-energy-level} and the complete results for all ions are listed in
\autoref{tab:reaction-scheme}.
\begin{table*}
  %\begin{tabular}{lp{7.5cm}cccc}
  %\begin{tabularx}{\textwidth}{llccXX}
  \begin{tabular*}{\textwidth}{@{\extracolsep{\fill}}lllllll}
    & \textbf{Reaction pathway} & \textbf{AE~\cite{Roeterdink:JCP117:6500}} & \textbf{Expected} &
    \multicolumn{2}{c}{\textbf{Experimental ionization order}} \\
    & \CFFII $\xrightarrow{\mathrm{Ionization/Excitation}}$& \textbf{(eV)} & \textbf{ionization order}& \textbf{Non-deflected} &
    \textbf{Deflected} \\[5pt]
    \arrayrulecolor{black}\hline
    R1 & \CFFIIion                              & 9.8   & 7 & 4.00(14) & 3.83(14)       \\[5pt]
    \arrayrulecolor{lightgray}\hline
    R2 & \CFFIion + \Iatom &10.3 & 7&\multirow{2}{*}{4.23(11)}&\multirow{2}{*}{4.20(12)}\\
    & \hspace{1em} $\xrightarrow{\mathrm{2nd~fragmentation}}$ \Iion / \CFFion& & & & \\[5pt]
    \arrayrulecolor{lightgray}\hline
    R3 & \IIion/\IIstarion +\CFF                & 10.5  & 7 & 4.13(13) & 4.06(9)        \\[5pt]
    \arrayrulecolor{lightgray}\hline
    R4 & \Iion +\CFFI & 12.7  & 8 &          &        \\
    R5 &  \IIstar + \CFF $\xrightarrow{\mathrm{Ionization}}$ \Iion + \Iatom +\CFF & 13.2 & 9 & 5.74(16) & 4.92(10) \\
    R6 & \CFFIstar + \Iatom $\xrightarrow{\mathrm{Ionization}}$ \Iion + \Iatom + \CFF
    & 13.2 & 9 & &  \\[5pt]
    \arrayrulecolor{lightgray}\hline
    R7 & \CFFion + \II  & 12.6 & 8 & \multirow{3}{*}{5.48(17)} & \multirow{3}{*}{4.78(9)} \\
    R8 & \CFFion + 2\Iatom & 14.2  & 9 &  &  \\
    & \hspace{1em}  $\xrightarrow{\mathrm{2nd~fragmentation}}$ \CFion & & & & \\[5pt]
    \arrayrulecolor{black} \hline
  \end{tabular*}
  \caption{Accessible reaction pathways following ionization of \CFFII for our experimental
     conditions, \ie, laser intensities. The appearance energy (AE) of the
     pathways~\cite{Roeterdink:JCP117:6500} is converted into the effective number of 800~nm
     (1.55~eV) photons and compared with the ionization orders estimated from the experiment.}
  \label{tab:reaction-scheme}
\end{table*}
The \CFFIIion, \CFFIion, and \IIion channels exhibit comparable ionization orders $\ordsim4$ at
both, the deflected and non-deflected positions. This indicates that these
ion channels are independent of the initial \CFFII rotational state. In contrast, the \Iion channel
exhibits a notable change in the ionization order between the two positions, decreasing from 5.73(16)
at the non-deflected position to 4.92(10) at the deflected position. A similar change is also observed
for the \CFFion channel.

\subsection{Rotational-state-dependent fragmentation}
\label{sec:rot_state_frag}
\begin{figure*}
   \centering \includegraphics[width=\linewidth]{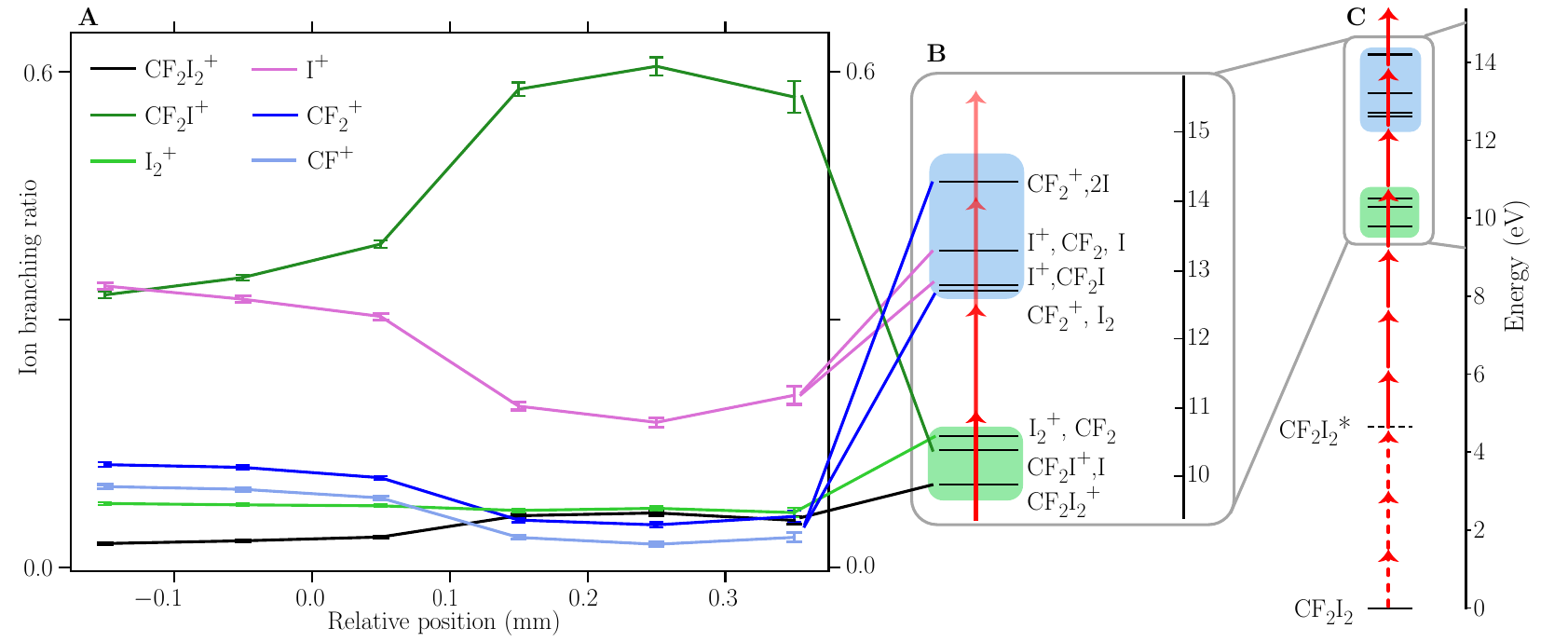}
   \caption{\textbf{A:} Measured branching ratios for the observed ion channels as a function of the
      relative position in the molecular beam. \textbf{B:} Assignment of the observed ion channels
      onto the low-energy (green) and high-energy (blue) bands. \textbf{C:} Simplified energy-level
      scheme with relevant states of \CFFII for our experiment. Vertical red arrows indicate the
      energy of one 800~nm (1.55~eV) photon. The dashed arrows denote transition via the
      intermediate resonant state \CFFIIast. \textbf{B} provides an expanded view of the ionic
      energy levels. The lines between panels \textbf{A} and \textbf{B} indicate the correspondence
      between each ion channel and its assigned energy band; see text for details.}
   \label{fig:summary-figure}
\end{figure*}

We observed pronounced variations in the branching ratios of the \CFFIIion fragmentation channels as
a function of the initial rotational-state distribution of \CFFII, which is correlated with the
position within the molecular beam profile. \autoref{fig:summary-figure} shows the branching ratios
measured at six positions spanning the region between the minimal and maximal deflection of the
molecular beam recorded in SMI mode.  At each position, the branching ratio is defined as the yield
of a specific fragment ion normalized to the total ion signal at a given vertical beam/ionization
position, providing a quantitative measure of the relative contributions.

The most pronounced changes are observed for the \CFFIion and \Iion channels. In the less-deflected
position, these channels exhibit comparable branching ratios of 0.330(3) and 0.341(1) for \CFFIion
and \Iion, respectively. Toward the strongly deflected position, the \CFFIion channel becomes
increasingly dominant, reaching a branching ratio of 0.569(18), while the contribution of the \Iion
channel correspondingly decreases to 0.208(11). Among the minor fragmentation channels, the
\CFFIIion contribution increases toward the deflected position, whereas the \CFFion and \CFion
contributions decrease. In contrast to all other channels, the \IIion channel exhibits only a weak
variation in branching ratio, changing from 0.077(1) in the non-deflected position to 0.066(5) in the
deflected position. This indicates that the \IIion branching ratio is largely independent of the
initial rotational state of \CFFII.

\section{Discussion}
\label{sec:Discussion}

The mass spectrum and fragmentation channels observed for \CFFII are qualitatively identical to
those found in dissociative ionization studies of \CHHII~\cite{Wang:IJQC106:1138,
   Zhang:JPB43:025102}, with the main channels \Iion and \CHHIion in this previous work
corresponding to channels \Iion and \CFFIion in the current work. The ionization order of \Iion
exceeds that of \CFFIion, consistent with an analogous situation for
\CHHII~\cite{Wang:IJQC106:1138}. Also, the minor channels \CFFIIion , \CFFion, and \CFion align with
those previously reported, replacing $\text{F}$ by $\text{H}$. However, there is a difference in the
behavior of the parent ion yield as a function of the laser intensity. The reduced \CHHII parent-ion
signal at high intensities ($\geq 100~\TWpcmcm$) was attributed to field-assisted
dissociation~\cite{Wang:IJQC106:1138}, where the strong laser field distorts cationic
potential-energy surfaces and drives fragmentation. In contrast, we observed a linear increase in
the \range{20}{50}~\TWpcmcm range, consistent with the expected intensity scaling under our
experimental conditions where the laser focus is much smaller than the molecular beam
diameter~\cite{Hankin:PRL84:5082, Hankin:PRA64:013405}. A further distinction is the laser
intensity, which was an order of magnitude higher in the \CHHII work~\cite{Wang:IJQC106:1138} than
in our present study. Despite this disparity, the overall fragmentation of \CFFII and \CHHII
behavior is qualitatively similar.

Analysis of the ionization-induced reaction pathways provides insight into the rotational-state
dependence of the ion product branching ratios. \autoref{tab:reaction-scheme} summarizes the
reaction pathways contributing to the individual fragment-ion channels accessible under our laser
conditions, along with the respective appearance energy
(AE)~\cite{Roeterdink:JCP117:6500,Farmanara:JCP113:1705,Radloff:CPL291:173}, the expected effective
number of photons required to induce each reaction path, and the experimentally determined
ionization orders at the deflected and non-deflected positions of the molecular beam. From the
comparison, we observe that the ionization-order numbers are significantly lower than the expected
ones. Earlier work demonstrated that single-photon excitation at 266~nm populates \CFFIIast with
dissociation lifetime on the order of $\ordsim100$~fs~\cite{Roeterdink:JCP117:6500,
  Radloff:CPL291:173}. Notably, the experimentally determined ionization orders are reduced by
approximately three 800~nm photons relative to the expected values. The three photon reduction
corresponds to the energy of a single 266~nm photon, providing evidence for resonance-enhanced
multi-photon ionization (REMPI) \emph{via} the intermediate \CFFIIast state with \oneBone symmetry,
consistent with previous studies~\citep{Roeterdink:JCP117:6500,Bergmann:JCP109:474}.

\autoref[(right half)]{fig:summary-figure} shows the energy-level diagram with the reaction pathways
and their respective AE according to \autoref{tab:reaction-scheme}. An individual red arrow
indicates absorption of a single 800~nm photon. The REMPI is indicated as a two-step process. The
first step of 3-photon absorption into the intermediate \CFFIIast state is marked by dashed red
arrows and the second step of multiphoton absorption into the respective ionic channel is marked by
solid red arrows.

Having established that multiple dissociation pathways are populated via REMPI, we now examine how
the initial rotational-state distribution governs the fragmentation of ionized \CFFII. We observe
systematic variations in the branching ratios across the spatially dispersed beam, corresponding to
decreasing lower rotational excitation of \CFFII. In particular, the significant increase in the
\CFFIion branching ratio is accompanied by a decrease in the \Iion branching ratio upon transition
to the rotationally colder ensemble. This indicates that the rotational-state distribution
influences the competition between fragmentation channels and alters charge localization during
fragmentation. The changes in the relative ion signals may arise either from selective population of
ionic states upon the strong-field interaction, influenced by rotational-state–dependent symmetry
constraints, or from fragmentation dynamics governed by energy thresholds associated with the
initial rotational excitation. It should be emphasized that the energy difference between the low
($\expectation{J}\approx2.70$) and high ($\expectation{J}\approx3.94$) rotational-state ensembles
corresponds to only a few~\ueV. In either scenario, the observed pronounced changes in the
fragmentation behavior are unexpected. While $M$-states could potentially also lead to similar
changes, we expect no strong $M$ asymmetry in our sample and therefore neglect it.

In the following discussion, possible reaction mechanisms are hypothesized based on our experimental
observations. The total ionization cross-section is assumed to be independent of the initial
rotational-state distribution, and the observed differences in branching ratios are attributed to
dynamics in the ionized state. This assumption is justified by the fact that the excitation energy
exceeds the ionization threshold by at least $\sim0.7$~eV. In addition, the differences in
appearance energies are of the same order of magnitude, such that both energy scales are large
compared to the rotational-energy scale, which is at the level of a few tens of \ueV. Consequently,
rotational selection rules are not expected to play a significant role in determining the total
ionization cross-section. This interpretation is further supported by the estimate that, at least,
several tens of rotational states are field-dressed and coupled at the applied laser intensities,
effectively eliminating rotational-state selectivity. Furthermore, the kicked rigid-rotor picture,
in which a rotational wave packet is created via Raman transitions that subsequently drive
fragmentation in a field-free regime, is assumed not to apply here. Estimates show that impulsive
alignment with a kick strength~\cite{Leibscher:PRL90:213001} of $P\approx20$ produces rotational
excitation involving on the order of 20~states~\cite{Karamatskos:NatComm10:3364}, largely
independent of the initial rotational-state distribution. Therefore, we attribute the strong
dependence on the initial rotational state to non-adiabatic dynamics during the laser pulse,
reflecting a breakdown of separability between rotational and vibrational motion, analogous to
Coriolis-type couplings. This interpretation motivates the consideration of rotationally mediated
couplings occurring during ion fragmentation. In \CFFIIion, strong spin–orbit interactions
associated with the iodine atoms, together with rotational motion, could promote coupling between
near-threshold ionic states and thereby alter the accessible fragmentation
pathways~\cite{Brown:JMS65:65}.

For the branching ratios, based on the appearance energies shown in \autoref{fig:summary-figure},
the \CFFIIion, \IIion, and \CFFIion channels are assigned to a low-energy band, while the \CFFion,
\CFion, and \Iion channels form a high-energy band, with these two bands being energetically
separated by more than one 800~nm photon. Ion channels in the low-energy band exhibit similar
ionization energies, whereas those in the high-energy band show changes in effective ionization
order between the rotationally warmer and colder molecular beams. For the low-energy channels, the
branching contributions of \CFFIIion and \CFFIion approximately double for the rotationally colder
ensemble, while the contribution of \IIion remains essentially constant. The \IIion can form either
\emph{via} detachment from \CFFIIion or \emph{via} dissociation of neutral \II from excited \CFFII,
followed by ionization~\cite{Roeterdink:JCP117:6500}; however, the latter pathway can be excluded
due to its longer timescale, $>500$~fs, compared to our laser-pulse duration. In contrast, the high-energy channels
   \Iion\ and \CFFion show reduced contributions and a decrease in ionization order by
approximately one photon when moving from the rotationally warmer to the colder ensemble. This is
counterintuitive, as rotational excitation would be expected to lower the required photon order.
Therefore, we infer that rotational excitation enhances either interband coupling, intraband
coupling within the high-energy band, or both. The dependence on the initial rotational distribution
further indicates that, despite similar excitation energies, the dynamics during the laser pulse are
governed by an increased accessible phase space in the laser-field-dressed manifold of coupled
states, leading to higher excitation probabilities and photon numbers. Overall, these observations
again support dynamics occurring in the presence of the laser field.

\begin{figure}
   \includegraphics[width=\linewidth]{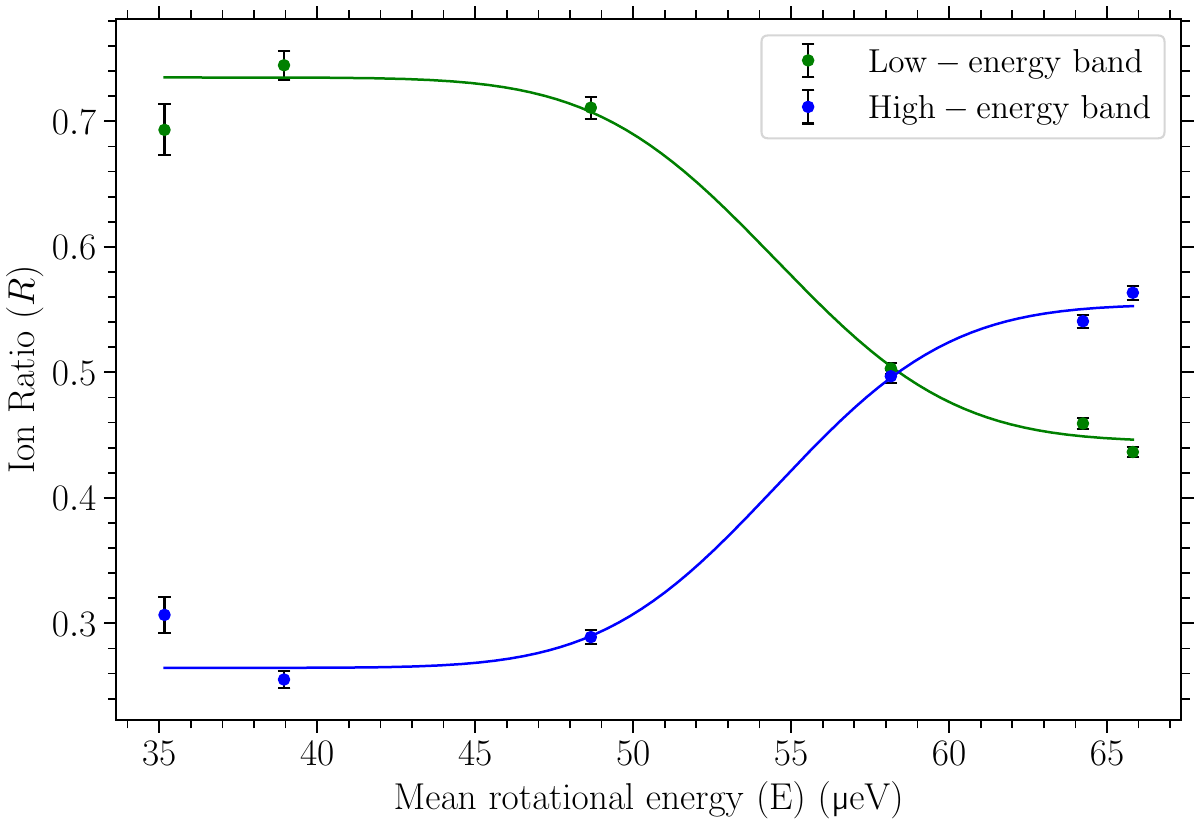}%
   \caption{Branching ratios of the low-energy and high-energy bands, cf.\
      \autoref{fig:summary-figure}, as a function of the mean rotational-state energy at a laser
      intensity of 31.3~\TWpcmcm. The measured dependencies were fitted with an error-function model
      to estimate the threshold rotational energy governing the redistribution between the two
      channels.}
   \label{fig:erf_func_model}
\end{figure}

To examine the role of rotational excitation, we analyzed the fragmentation dynamics as a function
of the mean rotational energy. The branching ratio between the low- and high-energy bands as a
function of mean rotational energy is shown in \autoref{fig:erf_func_model} as green and blue data
points, respectively.
This provides a direct measure of how the population shifts between the low- and
high-energy bands with increasing initial rotational energy. The high-energy band
population remains nearly constant at low rotational energies, then undergoes a threshold-like
increase before approaching a higher plateau at higher rotational energies. Since the fractional
change of the two channels is considered, the opposite
behavior applies to the low-energy band.

The dependence of the high-energy band branching ratio on the rotational
energy was fitted using an empirical error-function-based model,
\begin{equation*}
   R_{\text{H}}(E) = R_{\text{H,min}}  + \Delta R_{\text{H}}\frac{1}{2}\left[1 +
      \erf\left(\frac{E - E_0}{\sqrt{2}\sigma}\right)\right],
\end{equation*}

where $R_{\text{H}}$ is the branching ratio of the high-energy band, \ie, the sum of \CFFion, \CFion
and \Iion contributions. $E$ is the mean rotational energy, and $E_0$ is the central threshold
rotational energy. Further, $\Delta{}R_{\text{H}}=R_{\text{H,max}}-R_{\text{H,min}} $ denotes the
difference of the asymptotic branching ratios $ R_{\text{H,min}}$ and $R_{\text{H,max}}$ in the low
- and high-rotational-energy limits, respectively, while $\sigma$ characterizes the energy standard
deviation of the crossover region. For the lower band branching ratio, we have
$R_\text{L}=1-R_{\text{H}}$, which is equal to the contributions of \CFFIIion, \CFFIion, and \IIion.

From the fitting, shown as the solid blue line in \autoref{fig:erf_func_model}, we obtained a
threshold energy of $E_0 = 54.5(3)~\ueV$ and a standard deviation of $\sigma = 4.31(3)~\ueV$. The
population of the high-energy band asymptotically increases from $R_{\text{H,min}} = 0.26(1)$ to
$R_{\text{H,max}} = 0.55(1)$, and consequently, population of the low-energy band decreases from
$R_{\text{L,max}} = 0.73(1)$ to $R_{\text{L,min}} = 0.44(1)$. The small $\sigma$ is surprising given
the fact that the energy-sample standard deviation is in the order of $50~\ueV$ for all the
rotational-state distributions in the current experiment. This implies that a simple step-function
like behavior is not consistent with the data. Thus, we interpret a narrow energy range
$\Delta{}E=54.5~\pm4.31~\ueV$ centered around $E_0$ being relevant for the observed effects.
Accordingly, molecules with initial rotational energies in this range show enhanced excitation
probabilities to the upper band in the laser field. One way to interpret this, considering the
dissociating molecules, is that there may be a specific range of bond distances at which the
probability of excitation is increased. For molecules with higher initial rotational energy, which
should lead to faster dissociation, the time spent in the critical-distance range is smaller, and
thus the excitation probability is reduced. For molecules with small initial rotational energy, the
critical range is not reached at all during the laser pulse.

While our interpretation provides a fully plausible explanation of the clearly observed behavior, a
rigorous theoretical treatment remains beyond the scope of this manuscript, as a complete
description of the rovibronic structure of \CFFIIion is still challenging due to the interplay of
strong iodine- and fluorine-induced spin–orbit effects, with additional contributions from
rotationally induced coupling likely playing a role.

\section{Conclusion}

Overall, our experimental results demonstrate that the initial rotational-state distribution of
\CFFII, controlled \emph{via} electrostatic deflection, significantly modulates the probability of
strong-field dissociation pathways. To disentangle the underlying dissociation mechanisms, branching
ratios were analyzed in conjunction with a power dependence of the fragment-ion yields. The
power-dependence measurements reveal REMPI \emph{via} an
intermediate state, denoted as \CFFIIast.

We argue that the variation in the branching ratio arises from near-threshold ionic states, where a
narrow rotational energy gap steers the excitation in the ionic system. We attribute this to
non-adiabatic Coriolis-type coupling, which breaks the separability of rotational and vibrational
motion and thereby governs the competition between bound and dissociative continuum states. As a
result, the initial state preparation induces a selective redistribution of the product state.

Building on these findings, electron–ion coincidence measurements would enable a more complete
reconstruction of the ionization and fragmentation pathways. Pump–probe Coulomb-explosion
measurements could further resolve the timescale of the fragmentation dynamics. Complementary
theoretical studies are required to elucidate the interplay of rotational motion with spin–orbit,
vibronic, and Coriolis-type couplings.

In summary, precise knowledge of the initial state is crucial for accurate descriptions of
wavepacket preparation and subsequent photofragmentation dynamics. The pronounced dependence of the
post-ionization dynamics on the rotational quantum state further provides a means to influence
reaction outcomes--including charge localization and ion product branching ratios--through
classical, non-coherent-control-based rotational-state selection. By tuning the rotational energy on
the scale of only a few \ueV, the rotational state effectively serves as a finely tunable parameter
governing the accessible reaction pathways. These findings, in turn, motivate a discussion of how
control over the initial rotational-state distribution may be exploited to direct chemical
processes~\cite{Dantus:Science385:eadk1833, Zare:Science279:1875}.

\section*{Author contributions}
N.V.: Conceptualization, Data curation, Validation, Formal analysis, Investigation, Visualization,
Software (molecular trajectory simulations), Writing -- original draft.

I.S.V.: Investigation, Experiments, Validation, Writing -- review \& editing.

S.T.: Conceptualization, Methodology, Supervision, Visualization, Software ($J$-state coupling
simulations), Writing -- review \& editing.

J.K.: Conceptualization, Resources, Supervision, Funding acquisition, Writing -- review \& editing,
Project administration.

\section*{Conflicts of interest}
There are no conflicts to declare.

\section*{Data availability}
% Change the line when data is added to zenodo
The scripts used to analyze the recorded data and the specified equations are available from the
corresponding author upon request.

\section*{Acknowledgments}
We acknowledge financial support by Deutsches Elektronen-Synchrotron DESY, a member of the Helmholtz
Association (HGF), also for the provision of experimental facilities and for the use of the Maxwell
computational resources operated at DESY. The research was further supported by the Cluster of
Excellence ``CUI: Advanced Imaging of Matter'' of the Deutsche Forschungsgemeinschaft (DFG,
EXC~2056, project ID 390715994), the Helmholtz Foundation through funds from the Helmholtz-Lund
International Graduate School (HELIOS, HIRS-0018), the German Federal Ministry of Education and
Research (BMBF) and the Swedish Research Council (VR~2021-05992) through the Röntgen-Ångström
cluster project ``Ultrafast dynamics in intermolecular energy transfer: elementary processes in
aerosols and liquid chemistry'' (UDIET, 05K22GUA). The endstation for controlled-molecule
experiments (eCOMO) was funded with support from the Center for Molecular Water
Science. I.S.V.\ acknowledges support from the Alexander von Humboldt Foundation.

During the preparation of this work, ChatGPT was used for the purpose of language editing and
improving the clarity of the text. After using this service, the authors reviewed and edited the
content as needed and take full responsibility for the content of the publication.

\bibliography{string, cmi}

\begin{thebibliography}{60}%
\makeatletter
\providecommand \@ifxundefined [1]{%
 \@ifx{#1\undefined}
}%
\providecommand \@ifnum [1]{%
 \ifnum #1\expandafter \@firstoftwo
 \else \expandafter \@secondoftwo
 \fi
}%
\providecommand \@ifx [1]{%
 \ifx #1\expandafter \@firstoftwo
 \else \expandafter \@secondoftwo
 \fi
}%
\providecommand \natexlab [1]{#1}%
\providecommand \enquote  [1]{``#1''}%
\providecommand \bibnamefont  [1]{#1}%
\providecommand \bibfnamefont [1]{#1}%
\providecommand \citenamefont [1]{#1}%
\providecommand \href@noop [0]{\@secondoftwo}%
\providecommand \href [0]{\begingroup \@sanitize@url \@href}%
\providecommand \@href[1]{\@@startlink{#1}\@@href}%
\providecommand \@@href[1]{\endgroup#1\@@endlink}%
\providecommand \@sanitize@url [0]{\catcode `\\12\catcode `\$12\catcode
  `\&12\catcode `\#12\catcode `\^12\catcode `\_12\catcode `\%12\relax}%
\providecommand \@@startlink[1]{}%
\providecommand \@@endlink[0]{}%
\providecommand \url  [0]{\begingroup\@sanitize@url \@url }%
\providecommand \@url [1]{\endgroup\@href {#1}{\urlprefix }}%
\providecommand \urlprefix  [0]{URL }%
\providecommand \Eprint [0]{\href }%
\providecommand \doibase [0]{https://doi.org/}%
\providecommand \selectlanguage [0]{\@gobble}%
\providecommand \bibinfo  [0]{\@secondoftwo}%
\providecommand \bibfield  [0]{\@secondoftwo}%
\providecommand \translation [1]{[#1]}%
\providecommand \BibitemOpen [0]{}%
\providecommand \bibitemStop [0]{}%
\providecommand \bibitemNoStop [0]{.\EOS\space}%
\providecommand \EOS [0]{\spacefactor3000\relax}%
\providecommand \BibitemShut  [1]{\csname bibitem#1\endcsname}%
\let\auto@bib@innerbib\@empty
%</preamble>
\bibitem [{\citenamefont {Koenig}\ \emph {et~al.}(2021)\citenamefont {Koenig},
  \citenamefont {Volkamer}, \citenamefont {Apel}, \citenamefont {Bresch},
  \citenamefont {Cuevas}, \citenamefont {Dix}, \citenamefont {Eloranta},
  \citenamefont {Fernandez}, \citenamefont {Hall}, \citenamefont {Hornbrook},
  \citenamefont {Pierce}, \citenamefont {Reeves}, \citenamefont {Saiz-Lopez},\
  and\ \citenamefont {Ullmann}}]{Koenig:sciadv7:eabj6544}%
  \BibitemOpen
  \bibfield  {author} {\bibinfo {author} {\bibfnamefont {T.~K.}\ \bibnamefont
  {Koenig}}, \bibinfo {author} {\bibfnamefont {R.}~\bibnamefont {Volkamer}},
  \bibinfo {author} {\bibfnamefont {E.~C.}\ \bibnamefont {Apel}}, \bibinfo
  {author} {\bibfnamefont {J.~F.}\ \bibnamefont {Bresch}}, \bibinfo {author}
  {\bibfnamefont {C.~A.}\ \bibnamefont {Cuevas}}, \bibinfo {author}
  {\bibfnamefont {B.}~\bibnamefont {Dix}}, \bibinfo {author} {\bibfnamefont
  {E.~W.}\ \bibnamefont {Eloranta}}, \bibinfo {author} {\bibfnamefont {R.~P.}\
  \bibnamefont {Fernandez}}, \bibinfo {author} {\bibfnamefont {S.~R.}\
  \bibnamefont {Hall}}, \bibinfo {author} {\bibfnamefont {R.~S.}\ \bibnamefont
  {Hornbrook}}, \bibinfo {author} {\bibfnamefont {R.~B.}\ \bibnamefont
  {Pierce}}, \bibinfo {author} {\bibfnamefont {J.~M.}\ \bibnamefont {Reeves}},
  \bibinfo {author} {\bibfnamefont {A.}~\bibnamefont {Saiz-Lopez}}, and\
  \bibinfo {author} {\bibfnamefont {K.}~\bibnamefont {Ullmann}},\ }\bibfield
  {title} {\bibinfo {title} {Ozone depletion due to dust release of iodine in
  the free troposphere},\ }\href {https://doi.org/10.1126/sciadv.abj6544}
  {\bibfield  {journal} {\bibinfo  {journal} {Sci. Adv.}\ }\textbf {\bibinfo
  {volume} {7}},\ \bibinfo {pages} {eabj6544} (\bibinfo {year}
  {2021})}\BibitemShut {NoStop}%
\bibitem [{\citenamefont {O’Dowd}\ \emph {et~al.}(2002)\citenamefont
  {O’Dowd}, \citenamefont {Jimenez}, \citenamefont {Bahreini}, \citenamefont
  {Flagan}, \citenamefont {Seinfeld}, \citenamefont {H\"{a}meri}, \citenamefont
  {Pirjola}, \citenamefont {Kulmala}, \citenamefont {Jennings}, and\
  \citenamefont {Hoffmann}}]{ODowd:Nature417:632}%
  \BibitemOpen
  \bibfield  {author} {\bibinfo {author} {\bibfnamefont {C.~D.}\ \bibnamefont
  {O’Dowd}}, \bibinfo {author} {\bibfnamefont {J.~L.}\ \bibnamefont
  {Jimenez}}, \bibinfo {author} {\bibfnamefont {R.}~\bibnamefont {Bahreini}},
  \bibinfo {author} {\bibfnamefont {R.~C.}\ \bibnamefont {Flagan}}, \bibinfo
  {author} {\bibfnamefont {J.~H.}\ \bibnamefont {Seinfeld}}, \bibinfo {author}
  {\bibfnamefont {K.}~\bibnamefont {H\"{a}meri}}, \bibinfo {author}
  {\bibfnamefont {L.}~\bibnamefont {Pirjola}}, \bibinfo {author} {\bibfnamefont
  {M.}~\bibnamefont {Kulmala}}, \bibinfo {author} {\bibfnamefont {S.~G.}\
  \bibnamefont {Jennings}}, and\ \bibinfo {author} {\bibfnamefont
  {T.}~\bibnamefont {Hoffmann}},\ }\bibfield  {title} {\bibinfo {title} {Marine
  aerosol formation from biogenic iodine emissions},\ }\href
  {https://doi.org/10.1038/nature00775} {\bibfield  {journal} {\bibinfo
  {journal} {Nature}\ }\textbf {\bibinfo {volume} {417}},\ \bibinfo {pages}
  {632–636} (\bibinfo {year} {2002})}\BibitemShut {NoStop}%
\bibitem [{\citenamefont {Saiz-Lopez}\ \emph {et~al.}(2012)\citenamefont
  {Saiz-Lopez}, \citenamefont {Plane}, \citenamefont {Baker}, \citenamefont
  {Carpenter}, \citenamefont {von Glasow}, \citenamefont
  {G{\'{o}}mez~Mart{\'{i}}n}, \citenamefont {McFiggans}, and\ \citenamefont
  {Saunders}}]{Saiz-Lopez:CR112:1773}%
  \BibitemOpen
  \bibfield  {author} {\bibinfo {author} {\bibfnamefont {A.}~\bibnamefont
  {Saiz-Lopez}}, \bibinfo {author} {\bibfnamefont {J.~M.~C.}\ \bibnamefont
  {Plane}}, \bibinfo {author} {\bibfnamefont {A.~R.}\ \bibnamefont {Baker}},
  \bibinfo {author} {\bibfnamefont {L.~J.}\ \bibnamefont {Carpenter}}, \bibinfo
  {author} {\bibfnamefont {R.}~\bibnamefont {von Glasow}}, \bibinfo {author}
  {\bibfnamefont {J.~C.}\ \bibnamefont {G{\'{o}}mez~Mart{\'{i}}n}}, \bibinfo
  {author} {\bibfnamefont {G.}~\bibnamefont {McFiggans}}, and\ \bibinfo
  {author} {\bibfnamefont {R.~W.}\ \bibnamefont {Saunders}},\ }\bibfield
  {title} {\bibinfo {title} {Atmospheric chemistry of {I}odine},\ }\href
  {https://doi.org/10.1021/cr200029u} {\bibfield  {journal} {\bibinfo
  {journal} {Chem. Rev.}\ }\textbf {\bibinfo {volume} {112}},\ \bibinfo {pages}
  {1773} (\bibinfo {year} {2012})}\BibitemShut {NoStop}%
\bibitem [{\citenamefont {Gravestock}\ \emph {et~al.}(2010)\citenamefont
  {Gravestock}, \citenamefont {Blitz}, \citenamefont {Bloss}, and\
  \citenamefont {Heard}}]{Gravestock:CPC11:3928}%
  \BibitemOpen
  \bibfield  {author} {\bibinfo {author} {\bibfnamefont {T.~J.}\ \bibnamefont
  {Gravestock}}, \bibinfo {author} {\bibfnamefont {M.~A.}\ \bibnamefont
  {Blitz}}, \bibinfo {author} {\bibfnamefont {W.~J.}\ \bibnamefont {Bloss}},\
  and\ \bibinfo {author} {\bibfnamefont {D.~E.}\ \bibnamefont {Heard}},\
  }\bibfield  {title} {\bibinfo {title} {A multidimensional study of the
  reaction $\mathrm{CH_2I+O_2}$: Products and atmospheric implications},\
  }\href {https://doi.org/https://doi.org/10.1002/cphc.201000575} {\bibfield
  {journal} {\bibinfo  {journal} {Chem. Phys. Chem.}\ }\textbf {\bibinfo
  {volume} {11}},\ \bibinfo {pages} {3928} (\bibinfo {year}
  {2010})}\BibitemShut {NoStop}%
\bibitem [{\citenamefont {Suchan}\ \emph {et~al.}(2018)\citenamefont {Suchan},
  \citenamefont {Hollas}, \citenamefont {Curchod}, and\ \citenamefont
  {Slav{\'\i}{\v{c}}ek}}]{Suchan:FD212:307}%
  \BibitemOpen
  \bibfield  {author} {\bibinfo {author} {\bibfnamefont {J.}~\bibnamefont
  {Suchan}}, \bibinfo {author} {\bibfnamefont {D.}~\bibnamefont {Hollas}},
  \bibinfo {author} {\bibfnamefont {B.~F.~E.}\ \bibnamefont {Curchod}}, and\
  \bibinfo {author} {\bibfnamefont {P.}~\bibnamefont {Slav{\'\i}{\v{c}}ek}},\
  }\bibfield  {title} {\bibinfo {title} {On the importance of initial
  conditions for excited-state dynamics},\ }\href
  {https://doi.org/10.1039/c8fd00088c} {\bibfield  {journal} {\bibinfo
  {journal} {Faraday Disc.}\ }\textbf {\bibinfo {volume} {212}},\ \bibinfo
  {pages} {307} (\bibinfo {year} {2018})}\BibitemShut {NoStop}%
\bibitem [{\citenamefont {El-Khoury}\ \emph {et~al.}(2009)\citenamefont
  {El-Khoury}, \citenamefont {Tarnovsky}, \citenamefont {Schapiro},
  \citenamefont {Ryazantsev}, and\ \citenamefont
  {Olivucci}}]{El-Khoury:JPCA113:10767}%
  \BibitemOpen
  \bibfield  {author} {\bibinfo {author} {\bibfnamefont {P.~Z.}\ \bibnamefont
  {El-Khoury}}, \bibinfo {author} {\bibfnamefont {A.~N.}\ \bibnamefont
  {Tarnovsky}}, \bibinfo {author} {\bibfnamefont {I.}~\bibnamefont {Schapiro}},
  \bibinfo {author} {\bibfnamefont {M.~N.}\ \bibnamefont {Ryazantsev}}, and\
  \bibinfo {author} {\bibfnamefont {M.}~\bibnamefont {Olivucci}},\ }\bibfield
  {title} {\bibinfo {title} {Structure of the photochemical reaction path
  populated via promotion of $\text{CF}_2\text{I}_2$ into its first excited
  state},\ }\href {https://doi.org/10.1021/jp902873h} {\bibfield  {journal}
  {\bibinfo  {journal} {J. Phys. Chem. A}\ }\textbf {\bibinfo {volume} {113}},\
  \bibinfo {pages} {10767} (\bibinfo {year} {2009})}\BibitemShut {NoStop}%
\bibitem [{\citenamefont {Liu}\ \emph {et~al.}(2021)\citenamefont {Liu},
  \citenamefont {Li}, \citenamefont {Sun}, \citenamefont {Sun}, and\
  \citenamefont {Yang}}]{Liu:applsci11:1704}%
  \BibitemOpen
  \bibfield  {author} {\bibinfo {author} {\bibfnamefont {B.}~\bibnamefont
  {Liu}}, \bibinfo {author} {\bibfnamefont {Z.}~\bibnamefont {Li}}, \bibinfo
  {author} {\bibfnamefont {H.}~\bibnamefont {Sun}}, \bibinfo {author}
  {\bibfnamefont {Z.}~\bibnamefont {Sun}}, and\ \bibinfo {author}
  {\bibfnamefont {Y.}~\bibnamefont {Yang}},\ }\bibfield  {title} {\bibinfo
  {title} {Dissociative ionization of molecular {CF}$_2${B}r$_2$ under 800 and
  400~nm intense femtosecond laser fields},\ }\href
  {https://doi.org/10.3390/app11041704} {\bibfield  {journal} {\bibinfo
  {journal} {Appl. Sci.}\ }\textbf {\bibinfo {volume} {11}},\ \bibinfo {pages}
  {1704} (\bibinfo {year} {2021})}\BibitemShut {NoStop}%
\bibitem [{\citenamefont {Horton}\ \emph {et~al.}(2019)\citenamefont {Horton},
  \citenamefont {Liu}, \citenamefont {Forbes}, \citenamefont {Makhija},
  \citenamefont {Lausten}, \citenamefont {Stolow}, \citenamefont {Hockett},
  \citenamefont {Marquetand}, \citenamefont {Rozgonyi}, and\ \citenamefont
  {Weinacht}}]{Horton:JCP150:174201}%
  \BibitemOpen
  \bibfield  {author} {\bibinfo {author} {\bibfnamefont {S.~L.}\ \bibnamefont
  {Horton}}, \bibinfo {author} {\bibfnamefont {Y.}~\bibnamefont {Liu}},
  \bibinfo {author} {\bibfnamefont {R.}~\bibnamefont {Forbes}}, \bibinfo
  {author} {\bibfnamefont {V.}~\bibnamefont {Makhija}}, \bibinfo {author}
  {\bibfnamefont {R.}~\bibnamefont {Lausten}}, \bibinfo {author} {\bibfnamefont
  {A.}~\bibnamefont {Stolow}}, \bibinfo {author} {\bibfnamefont
  {P.}~\bibnamefont {Hockett}}, \bibinfo {author} {\bibfnamefont
  {P.}~\bibnamefont {Marquetand}}, \bibinfo {author} {\bibfnamefont
  {T.}~\bibnamefont {Rozgonyi}}, and\ \bibinfo {author} {\bibfnamefont
  {T.}~\bibnamefont {Weinacht}},\ }\bibfield  {title} {\bibinfo {title}
  {Excited state dynamics of $\mathrm{CH_2I_2}$ and $\mathrm{CH_2BrI}$ studied
  with {UV} pump {VUV} probe photoelectron spectroscopy},\ }\href
  {https://doi.org/10.1063/1.5086665} {\bibfield  {journal} {\bibinfo
  {journal} {J. Chem. Phys.}\ }\textbf {\bibinfo {volume} {150}},\ \bibinfo
  {pages} {174201} (\bibinfo {year} {2019})}\BibitemShut {NoStop}%
\bibitem [{\citenamefont {Recio}\ \emph {et~al.}(2022)\citenamefont {Recio},
  \citenamefont {Cachón}, \citenamefont {Rubio-Lago}, \citenamefont
  {Chicharro}, \citenamefont {Zanchet}, \citenamefont {Limão-Vieira},
  \citenamefont {de~Oliveira}, \citenamefont {Samartzis}, \citenamefont
  {Marggi~Poullain}, and\ \citenamefont {Ba{\~n}ares}}]{Recio:JPCA126:8404}%
  \BibitemOpen
  \bibfield  {author} {\bibinfo {author} {\bibfnamefont {P.}~\bibnamefont
  {Recio}}, \bibinfo {author} {\bibfnamefont {J.}~\bibnamefont {Cachón}},
  \bibinfo {author} {\bibfnamefont {L.}~\bibnamefont {Rubio-Lago}}, \bibinfo
  {author} {\bibfnamefont {D.~V.}\ \bibnamefont {Chicharro}}, \bibinfo {author}
  {\bibfnamefont {A.}~\bibnamefont {Zanchet}}, \bibinfo {author} {\bibfnamefont
  {P.}~\bibnamefont {Limão-Vieira}}, \bibinfo {author} {\bibfnamefont
  {N.}~\bibnamefont {de~Oliveira}}, \bibinfo {author} {\bibfnamefont {P.~C.}\
  \bibnamefont {Samartzis}}, \bibinfo {author} {\bibfnamefont {S.}~\bibnamefont
  {Marggi~Poullain}}, and\ \bibinfo {author} {\bibfnamefont {L.}~\bibnamefont
  {Ba{\~n}ares}},\ }\bibfield  {title} {\bibinfo {title} {Imaging the
  photodissociation dynamics and fragment alignment of $\mathrm{CH_2BrI}$ at
  193 nm},\ }\href {https://doi.org/10.1021/acs.jpca.2c05897} {\bibfield
  {journal} {\bibinfo  {journal} {J. Phys. Chem. A}\ }\textbf {\bibinfo
  {volume} {126}},\ \bibinfo {pages} {8404} (\bibinfo {year}
  {2022})}\BibitemShut {NoStop}%
\bibitem [{\citenamefont {Wannenmacher}\ \emph {et~al.}(1991)\citenamefont
  {Wannenmacher}, \citenamefont {Felder}, and\ \citenamefont
  {Huber}}]{Wannenmacher:JCP95:986}%
  \BibitemOpen
  \bibfield  {author} {\bibinfo {author} {\bibfnamefont {E.~A.~J.}\
  \bibnamefont {Wannenmacher}}, \bibinfo {author} {\bibfnamefont
  {P.}~\bibnamefont {Felder}}, and\ \bibinfo {author} {\bibfnamefont {J.~R.}\
  \bibnamefont {Huber}},\ }\bibfield  {title} {\bibinfo {title} {The
  simultaneous three‐body dissociation of $\mathrm{CF_2I_2}$},\ }\href
  {https://doi.org/10.1063/1.461054} {\bibfield  {journal} {\bibinfo  {journal}
  {J. Chem. Phys.}\ }\textbf {\bibinfo {volume} {95}},\ \bibinfo {pages} {986}
  (\bibinfo {year} {1991})}\BibitemShut {NoStop}%
\bibitem [{\citenamefont {Toulson}\ \emph {et~al.}(2016)\citenamefont
  {Toulson}, \citenamefont {Alaniz}, \citenamefont {Hill}, and\ \citenamefont
  {Murray}}]{Toulson:PCCP18:11091}%
  \BibitemOpen
  \bibfield  {author} {\bibinfo {author} {\bibfnamefont {B.~W.}\ \bibnamefont
  {Toulson}}, \bibinfo {author} {\bibfnamefont {J.~P.}\ \bibnamefont {Alaniz}},
  \bibinfo {author} {\bibfnamefont {J.~G.}\ \bibnamefont {Hill}}, and\
  \bibinfo {author} {\bibfnamefont {C.}~\bibnamefont {Murray}},\ }\bibfield
  {title} {\bibinfo {title} {Near-{UV} photodissociation dynamics of
  $\mathrm{CH_2I_2}$},\ }\href {https://doi.org/10.1039/c6cp01063f} {\bibfield
  {journal} {\bibinfo  {journal} {Phys. Chem. Chem. Phys.}\ }\textbf {\bibinfo
  {volume} {18}},\ \bibinfo {pages} {11091} (\bibinfo {year}
  {2016})}\BibitemShut {NoStop}%
\bibitem [{\citenamefont {Baum}\ \emph {et~al.}(1993)\citenamefont {Baum},
  \citenamefont {Felder}, and\ \citenamefont {Huber}}]{Baum:JCP98:1999}%
  \BibitemOpen
  \bibfield  {author} {\bibinfo {author} {\bibfnamefont {G.}~\bibnamefont
  {Baum}}, \bibinfo {author} {\bibfnamefont {P.}~\bibnamefont {Felder}}, and\
  \bibinfo {author} {\bibfnamefont {J.~R.}\ \bibnamefont {Huber}},\ }\bibfield
  {title} {\bibinfo {title} {Photofragmentation of $\mathrm{CF_2I_2}$.
  {C}ompetition between radical and three‐body dissociation},\ }\href
  {https://doi.org/10.1063/1.464233} {\bibfield  {journal} {\bibinfo  {journal}
  {J. Chem. Phys.}\ }\textbf {\bibinfo {volume} {98}},\ \bibinfo {pages} {1999}
  (\bibinfo {year} {1993})}\BibitemShut {NoStop}%
\bibitem [{\citenamefont {Bergmann}\ \emph {et~al.}(1998)\citenamefont
  {Bergmann}, \citenamefont {Carter}, \citenamefont {Hall}, and\ \citenamefont
  {Huber}}]{Bergmann:JCP109:474}%
  \BibitemOpen
  \bibfield  {author} {\bibinfo {author} {\bibfnamefont {K.}~\bibnamefont
  {Bergmann}}, \bibinfo {author} {\bibfnamefont {R.~T.}\ \bibnamefont
  {Carter}}, \bibinfo {author} {\bibfnamefont {G.~E.}\ \bibnamefont {Hall}},\
  and\ \bibinfo {author} {\bibfnamefont {J.~R.}\ \bibnamefont {Huber}},\
  }\bibfield  {title} {\bibinfo {title} {Resonance enhanced multiphoton
  ionization time-of-flight study of $\mathrm{CF_2I_2}$ photodissociation},\
  }\href {https://doi.org/10.1063/1.476670} {\bibfield  {journal} {\bibinfo
  {journal} {J. Chem. Phys.}\ }\textbf {\bibinfo {volume} {109}},\ \bibinfo
  {pages} {474} (\bibinfo {year} {1998})}\BibitemShut {NoStop}%
\bibitem [{\citenamefont {Radloff}\ \emph {et~al.}(1998)\citenamefont
  {Radloff}, \citenamefont {Farmanara}, \citenamefont {Stert}, \citenamefont
  {Schreiber}, and\ \citenamefont {Huber}}]{Radloff:CPL291:173}%
  \BibitemOpen
  \bibfield  {author} {\bibinfo {author} {\bibfnamefont {W.}~\bibnamefont
  {Radloff}}, \bibinfo {author} {\bibfnamefont {P.}~\bibnamefont {Farmanara}},
  \bibinfo {author} {\bibfnamefont {V.}~\bibnamefont {Stert}}, \bibinfo
  {author} {\bibfnamefont {E.}~\bibnamefont {Schreiber}}, and\ \bibinfo
  {author} {\bibfnamefont {J.}~\bibnamefont {Huber}},\ }\bibfield  {title}
  {\bibinfo {title} {Ultrafast photodissociation dynamics of electronically
  excited $\mathrm{CF_2I_2}$ molecules},\ }\href
  {https://doi.org/10.1016/s0009-2614(98)00551-x} {\bibfield  {journal}
  {\bibinfo  {journal} {Chem. Phys. Lett.}\ }\textbf {\bibinfo {volume}
  {291}},\ \bibinfo {pages} {173} (\bibinfo {year} {1998})}\BibitemShut
  {NoStop}%
\bibitem [{\citenamefont {Farmanara}\ \emph {et~al.}(2000)\citenamefont
  {Farmanara}, \citenamefont {Stert}, \citenamefont {Ritze}, and\
  \citenamefont {Radloff}}]{Farmanara:JCP113:1705}%
  \BibitemOpen
  \bibfield  {author} {\bibinfo {author} {\bibfnamefont {P.}~\bibnamefont
  {Farmanara}}, \bibinfo {author} {\bibfnamefont {V.}~\bibnamefont {Stert}},
  \bibinfo {author} {\bibfnamefont {H.-H.}\ \bibnamefont {Ritze}}, and\
  \bibinfo {author} {\bibfnamefont {W.}~\bibnamefont {Radloff}},\ }\bibfield
  {title} {\bibinfo {title} {Analysis of the ultrafast photodissociation of
  electronically excited $\mathrm{CF_2I_2}$ molecules by femtosecond
  time-resolved photoelectron spectroscopy},\ }\href
  {https://doi.org/10.1063/1.481972} {\bibfield  {journal} {\bibinfo  {journal}
  {J. Chem. Phys.}\ }\textbf {\bibinfo {volume} {113}},\ \bibinfo {pages}
  {1705} (\bibinfo {year} {2000})}\BibitemShut {NoStop}%
\bibitem [{\citenamefont {Roeterdink} and\ \citenamefont
  {Janssen}(2002)}]{Roeterdink:JCP117:6500}%
  \BibitemOpen
  \bibfield  {author} {\bibinfo {author} {\bibfnamefont {W.~G.}\ \bibnamefont
  {Roeterdink}} and\ \bibinfo {author} {\bibfnamefont {M.~H.~M.}\ \bibnamefont
  {Janssen}},\ }\bibfield  {title} {\bibinfo {title} {Femtosecond velocity map
  imaging of concerted photodynamics in $\mathrm{CF_2I_2}$},\ }\href
  {https://doi.org/10.1063/1.1505026} {\bibfield  {journal} {\bibinfo
  {journal} {J. Chem. Phys.}\ }\textbf {\bibinfo {volume} {117}},\ \bibinfo
  {pages} {6500} (\bibinfo {year} {2002})}\BibitemShut {NoStop}%
\bibitem [{\citenamefont {Townsend}\ \emph {et~al.}(2004)\citenamefont
  {Townsend}, \citenamefont {Lahankar}, \citenamefont {Lee}, \citenamefont
  {Chambreau}, \citenamefont {Suits}, \citenamefont {Zhang}, \citenamefont
  {Rheinecker}, \citenamefont {Harding}, and\ \citenamefont
  {Bowman}}]{Townsend:Science306:1158}%
  \BibitemOpen
  \bibfield  {author} {\bibinfo {author} {\bibfnamefont {D.}~\bibnamefont
  {Townsend}}, \bibinfo {author} {\bibfnamefont {S.~A.}\ \bibnamefont
  {Lahankar}}, \bibinfo {author} {\bibfnamefont {S.~K.}\ \bibnamefont {Lee}},
  \bibinfo {author} {\bibfnamefont {S.~D.}\ \bibnamefont {Chambreau}}, \bibinfo
  {author} {\bibfnamefont {A.~G.}\ \bibnamefont {Suits}}, \bibinfo {author}
  {\bibfnamefont {X.}~\bibnamefont {Zhang}}, \bibinfo {author} {\bibfnamefont
  {J.}~\bibnamefont {Rheinecker}}, \bibinfo {author} {\bibfnamefont {L.~B.}\
  \bibnamefont {Harding}}, and\ \bibinfo {author} {\bibfnamefont {J.~M.}\
  \bibnamefont {Bowman}},\ }\bibfield  {title} {\bibinfo {title} {The roaming
  atom: Straying from the reaction path in formaldehyde decomposition},\ }\href
  {https://doi.org/10.1126/science.1104386} {\bibfield  {journal} {\bibinfo
  {journal} {Science}\ }\textbf {\bibinfo {volume} {306}},\ \bibinfo {pages}
  {1158} (\bibinfo {year} {2004})}\BibitemShut {NoStop}%
\bibitem [{\citenamefont {Borin}\ \emph {et~al.}(2016)\citenamefont {Borin},
  \citenamefont {Matveev}, \citenamefont {Budkina}, \citenamefont {El-Khoury},\
  and\ \citenamefont {Tarnovsky}}]{Borin:PCCP18:28883}%
  \BibitemOpen
  \bibfield  {author} {\bibinfo {author} {\bibfnamefont {V.~A.}\ \bibnamefont
  {Borin}}, \bibinfo {author} {\bibfnamefont {S.~M.}\ \bibnamefont {Matveev}},
  \bibinfo {author} {\bibfnamefont {D.~S.}\ \bibnamefont {Budkina}}, \bibinfo
  {author} {\bibfnamefont {P.~Z.}\ \bibnamefont {El-Khoury}}, and\ \bibinfo
  {author} {\bibfnamefont {A.~N.}\ \bibnamefont {Tarnovsky}},\ }\bibfield
  {title} {\bibinfo {title} {{Direct photoisomerization of
  $\text{CH}_2\text{I}_2$ \textit{vs.} $\text{CHBr}_3$ in the gas phase: a
  joint 50~fs experimental and multireference resonance-theoretical study}},\
  }\href {https://doi.org/10.1039/c6cp05129d} {\bibfield  {journal} {\bibinfo
  {journal} {Phys. Chem. Chem. Phys.}\ }\textbf {\bibinfo {volume} {18}},\
  \bibinfo {pages} {28883} (\bibinfo {year} {2016})}\BibitemShut {NoStop}%
\bibitem [{\citenamefont {Anderson}\ \emph {et~al.}(2013)\citenamefont
  {Anderson}, \citenamefont {Spears}, \citenamefont {Wilson}, and\
  \citenamefont {Sension}}]{Anderson:JCP139:194307}%
  \BibitemOpen
  \bibfield  {author} {\bibinfo {author} {\bibfnamefont {C.~P.}\ \bibnamefont
  {Anderson}}, \bibinfo {author} {\bibfnamefont {K.~G.}\ \bibnamefont
  {Spears}}, \bibinfo {author} {\bibfnamefont {K.~R.}\ \bibnamefont {Wilson}},\
  and\ \bibinfo {author} {\bibfnamefont {R.~J.}\ \bibnamefont {Sension}},\
  }\bibfield  {title} {\bibinfo {title} {Solvent dependent branching between
  $\mathrm{C-I}$ and $\mathrm{C-Br}$ bond cleavage following 266 nm excitation
  of $\mathrm{CH_2BrI}$},\ }\href {https://doi.org/10.1063/1.4829899}
  {\bibfield  {journal} {\bibinfo  {journal} {J. Chem. Phys.}\ }\textbf
  {\bibinfo {volume} {139}},\ \bibinfo {pages} {194307} (\bibinfo {year}
  {2013})}\BibitemShut {NoStop}%
\bibitem [{\citenamefont {El-Khoury}\ \emph {et~al.}(2010)\citenamefont
  {El-Khoury}, \citenamefont {George}, \citenamefont {Kalume}, \citenamefont
  {Reid}, \citenamefont {Ault}, and\ \citenamefont
  {Tarnovsky}}]{El-Khoury:JCP132:124501}%
  \BibitemOpen
  \bibfield  {author} {\bibinfo {author} {\bibfnamefont {P.~Z.}\ \bibnamefont
  {El-Khoury}}, \bibinfo {author} {\bibfnamefont {L.}~\bibnamefont {George}},
  \bibinfo {author} {\bibfnamefont {A.}~\bibnamefont {Kalume}}, \bibinfo
  {author} {\bibfnamefont {S.~A.}\ \bibnamefont {Reid}}, \bibinfo {author}
  {\bibfnamefont {B.~S.}\ \bibnamefont {Ault}}, and\ \bibinfo {author}
  {\bibfnamefont {A.~N.}\ \bibnamefont {Tarnovsky}},\ }\bibfield  {title}
  {\bibinfo {title} {Characterization of $iso-\mathrm{CF_2I_2}$ in frequency
  and ultrafast time domains},\ }\href {https://doi.org/10.1063/1.3357728}
  {\bibfield  {journal} {\bibinfo  {journal} {J. Chem. Phys.}\ }\textbf
  {\bibinfo {volume} {132}},\ \bibinfo {pages} {124501} (\bibinfo {year}
  {2010})}\BibitemShut {NoStop}%
\bibitem [{\citenamefont {Pearson}\ \emph {et~al.}(2007)\citenamefont
  {Pearson}, \citenamefont {Nichols}, and\ \citenamefont
  {Weinacht}}]{Pearson:JCP127:131101}%
  \BibitemOpen
  \bibfield  {author} {\bibinfo {author} {\bibfnamefont {B.~J.}\ \bibnamefont
  {Pearson}}, \bibinfo {author} {\bibfnamefont {S.~R.}\ \bibnamefont
  {Nichols}}, and\ \bibinfo {author} {\bibfnamefont {T.}~\bibnamefont
  {Weinacht}},\ }\bibfield  {title} {\bibinfo {title} {Molecular fragmentation
  driven by ultrafast dynamic ionic resonances},\ }\href
  {https://doi.org/10.1063/1.2790419} {\bibfield  {journal} {\bibinfo
  {journal} {J. Chem. Phys.}\ }\textbf {\bibinfo {volume} {127}},\ \bibinfo
  {pages} {131101} (\bibinfo {year} {2007})}\BibitemShut {NoStop}%
\bibitem [{\citenamefont {Crim}(1993)}]{Crim:ARPC1:397}%
  \BibitemOpen
  \bibfield  {author} {\bibinfo {author} {\bibfnamefont {F.~F.}\ \bibnamefont
  {Crim}},\ }\bibfield  {title} {\bibinfo {title} {Vibrationally mediated
  photodissociation: Exploring excited-state surfaces and controlling
  decomposition pathways},\ }\href
  {https://doi.org/10.1146/annurev.pc.44.100193.002145} {\bibfield  {journal}
  {\bibinfo  {journal} {Annu. Rev. Phys. Chem.}\ }\textbf {\bibinfo {volume}
  {44}},\ \bibinfo {pages} {397} (\bibinfo {year} {1993})}\BibitemShut
  {NoStop}%
\bibitem [{\citenamefont {Prlj}\ \emph {et~al.}(2023)\citenamefont {Prlj},
  \citenamefont {Hollas}, and\ \citenamefont {Curchod}}]{Prlj:JPCA127:7400}%
  \BibitemOpen
  \bibfield  {author} {\bibinfo {author} {\bibfnamefont {A.}~\bibnamefont
  {Prlj}}, \bibinfo {author} {\bibfnamefont {D.}~\bibnamefont {Hollas}}, and\
  \bibinfo {author} {\bibfnamefont {B.~F.~E.}\ \bibnamefont {Curchod}},\
  }\bibfield  {title} {\bibinfo {title} {Deciphering the influence of
  ground-state distributions on the calculation of photolysis observables},\
  }\href {https://doi.org/10.1021/acs.jpca.3c02333} {\bibfield  {journal}
  {\bibinfo  {journal} {J. Phys. Chem. A}\ }\textbf {\bibinfo {volume} {127}},\
  \bibinfo {pages} {7400} (\bibinfo {year} {2023})}\BibitemShut {NoStop}%
\bibitem [{\citenamefont {Bamford}\ \emph {et~al.}(1985)\citenamefont
  {Bamford}, \citenamefont {Filseth}, \citenamefont {Foltz}, \citenamefont
  {Hepburn}, and\ \citenamefont {Moore}}]{Bamford:JCP82:3032}%
  \BibitemOpen
  \bibfield  {author} {\bibinfo {author} {\bibfnamefont {D.~J.}\ \bibnamefont
  {Bamford}}, \bibinfo {author} {\bibfnamefont {S.~V.}\ \bibnamefont
  {Filseth}}, \bibinfo {author} {\bibfnamefont {M.~F.}\ \bibnamefont {Foltz}},
  \bibinfo {author} {\bibfnamefont {J.~W.}\ \bibnamefont {Hepburn}}, and\
  \bibinfo {author} {\bibfnamefont {C.~B.}\ \bibnamefont {Moore}},\ }\bibfield
  {title} {\bibinfo {title} {Photofragmentation dynamics of formaldehyde:
  {CO}($\nu, {J}$) distributions as a function of initial rovibronic state and
  isotopic substitution},\ }\href {https://doi.org/10.1063/1.448252} {\bibfield
   {journal} {\bibinfo  {journal} {J. Chem. Phys.}\ }\textbf {\bibinfo {volume}
  {82}},\ \bibinfo {pages} {3032–3041} (\bibinfo {year} {1985})}\BibitemShut
  {NoStop}%
\bibitem [{\citenamefont {Xie}\ \emph {et~al.}(2000)\citenamefont {Xie},
  \citenamefont {Guo}, \citenamefont {Amatatsu}, and\ \citenamefont
  {Kosloff}}]{Xie:JPCA104:1009}%
  \BibitemOpen
  \bibfield  {author} {\bibinfo {author} {\bibfnamefont {D.}~\bibnamefont
  {Xie}}, \bibinfo {author} {\bibfnamefont {H.}~\bibnamefont {Guo}}, \bibinfo
  {author} {\bibfnamefont {Y.}~\bibnamefont {Amatatsu}}, and\ \bibinfo
  {author} {\bibfnamefont {R.}~\bibnamefont {Kosloff}},\ }\bibfield  {title}
  {\bibinfo {title} {Three-dimensional photodissociation dynamics of rotational
  state selected methyl iodide},\ }\href {https://doi.org/10.1021/jp9932463}
  {\bibfield  {journal} {\bibinfo  {journal} {J. Phys. Chem. A}\ }\textbf
  {\bibinfo {volume} {104}},\ \bibinfo {pages} {1009} (\bibinfo {year}
  {2000})}\BibitemShut {NoStop}%
\bibitem [{\citenamefont {Jano{\v{s}}}\ \emph {et~al.}(2025)\citenamefont
  {Jano{\v{s}}}, \citenamefont {Slav{\'\i}{\v{c}}ek}, and\ \citenamefont
  {Curchod}}]{Janos:ACR58:261}%
  \BibitemOpen
  \bibfield  {author} {\bibinfo {author} {\bibfnamefont {J.}~\bibnamefont
  {Jano{\v{s}}}}, \bibinfo {author} {\bibfnamefont {P.}~\bibnamefont
  {Slav{\'\i}{\v{c}}ek}}, and\ \bibinfo {author} {\bibfnamefont {B.~F.~E.}\
  \bibnamefont {Curchod}},\ }\bibfield  {title} {\bibinfo {title} {{Selecting
  Initial Conditions for Trajectory-Based Nonadiabatic Simulations}},\ }\href
  {https://doi.org/10.1021/acs.accounts.4c00687} {\bibfield  {journal}
  {\bibinfo  {journal} {Acc. Chem. Res.}\ }\textbf {\bibinfo {volume} {58}},\
  \bibinfo {pages} {261} (\bibinfo {year} {2025})}\BibitemShut {NoStop}%
\bibitem [{\citenamefont {Vinkl{\'a}rek}\ \emph {et~al.}(2021)\citenamefont
  {Vinkl{\'a}rek}, \citenamefont {Suchan}, \citenamefont {Rakovsk{\`y}},
  \citenamefont {Moriov{\'a}}, \citenamefont {Poterya}, \citenamefont
  {Slav{\'\i}{\v{c}}ek}, and\ \citenamefont
  {F{\'a}rn{\'\i}k}}]{Vinklarek:PCCP23:14340}%
  \BibitemOpen
  \bibfield  {author} {\bibinfo {author} {\bibfnamefont {I.~S.}\ \bibnamefont
  {Vinkl{\'a}rek}}, \bibinfo {author} {\bibfnamefont {J.}~\bibnamefont
  {Suchan}}, \bibinfo {author} {\bibfnamefont {J.}~\bibnamefont
  {Rakovsk{\`y}}}, \bibinfo {author} {\bibfnamefont {K.}~\bibnamefont
  {Moriov{\'a}}}, \bibinfo {author} {\bibfnamefont {V.}~\bibnamefont
  {Poterya}}, \bibinfo {author} {\bibfnamefont {P.}~\bibnamefont
  {Slav{\'\i}{\v{c}}ek}}, and\ \bibinfo {author} {\bibfnamefont
  {M.}~\bibnamefont {F{\'a}rn{\'\i}k}},\ }\bibfield  {title} {\bibinfo {title}
  {Energy partitioning and spin--orbit effects in the photodissociation of
  higher chloroalkanes},\ }\href {https://doi.org/10.1039/D1CP01371H}
  {\bibfield  {journal} {\bibinfo  {journal} {Phys. Chem. Chem. Phys.}\
  }\textbf {\bibinfo {volume} {23}},\ \bibinfo {pages} {14340} (\bibinfo {year}
  {2021})}\BibitemShut {NoStop}%
\bibitem [{\citenamefont {Murillo-Sánchez}\ \emph {et~al.}(2018)\citenamefont
  {Murillo-Sánchez}, \citenamefont {Marggi~Poullain}, \citenamefont {Bajo},
  \citenamefont {Corrales}, \citenamefont {González-Vázquez}, \citenamefont
  {Solá}, and\ \citenamefont {Bañares}}]{MurilloSanchez:PCCP20:20766}%
  \BibitemOpen
  \bibfield  {author} {\bibinfo {author} {\bibfnamefont {M.~L.}\ \bibnamefont
  {Murillo-Sánchez}}, \bibinfo {author} {\bibfnamefont {S.}~\bibnamefont
  {Marggi~Poullain}}, \bibinfo {author} {\bibfnamefont {J.~J.}\ \bibnamefont
  {Bajo}}, \bibinfo {author} {\bibfnamefont {M.~E.}\ \bibnamefont {Corrales}},
  \bibinfo {author} {\bibfnamefont {J.}~\bibnamefont {González-Vázquez}},
  \bibinfo {author} {\bibfnamefont {I.~R.}\ \bibnamefont {Solá}}, and\
  \bibinfo {author} {\bibfnamefont {L.}~\bibnamefont {Bañares}},\ }\bibfield
  {title} {\bibinfo {title} {Halogen-atom effect on the ultrafast
  photodissociation dynamics of the dihalomethanes $\mathrm{CH_2ICl}$ and
  $\mathrm{CH_2BrI}$},\ }\href {https://doi.org/10.1039/c8cp03600d} {\bibfield
  {journal} {\bibinfo  {journal} {Phys. Chem. Chem. Phys.}\ }\textbf {\bibinfo
  {volume} {20}},\ \bibinfo {pages} {20766–20778} (\bibinfo {year}
  {2018})}\BibitemShut {NoStop}%
\bibitem [{\citenamefont {Trost}\ \emph {et~al.}(2025)\citenamefont {Trost},
  \citenamefont {Díaz-Tendero}, \citenamefont {Lindenblatt}, \citenamefont
  {Meister}, \citenamefont {Schnorr}, \citenamefont {Augustin}, \citenamefont
  {Schmid}, \citenamefont {Liu}, \citenamefont {Schoch}, \citenamefont
  {Hosseini}, \citenamefont {Zmerli}, \citenamefont {Guillemin}, \citenamefont
  {Piancastelli}, \citenamefont {Braune}, \citenamefont {Schr\"{o}ter},
  \citenamefont {Pfeifer}, \citenamefont {Martín}, \citenamefont {Simon},\
  and\ \citenamefont {Moshammer}}]{Trost:jpbamop58:085101}%
  \BibitemOpen
  \bibfield  {author} {\bibinfo {author} {\bibfnamefont {F.}~\bibnamefont
  {Trost}}, \bibinfo {author} {\bibfnamefont {S.}~\bibnamefont
  {Díaz-Tendero}}, \bibinfo {author} {\bibfnamefont {H.}~\bibnamefont
  {Lindenblatt}}, \bibinfo {author} {\bibfnamefont {S.}~\bibnamefont
  {Meister}}, \bibinfo {author} {\bibfnamefont {K.}~\bibnamefont {Schnorr}},
  \bibinfo {author} {\bibfnamefont {S.}~\bibnamefont {Augustin}}, \bibinfo
  {author} {\bibfnamefont {G.}~\bibnamefont {Schmid}}, \bibinfo {author}
  {\bibfnamefont {Y.}~\bibnamefont {Liu}}, \bibinfo {author} {\bibfnamefont
  {P.}~\bibnamefont {Schoch}}, \bibinfo {author} {\bibfnamefont
  {F.}~\bibnamefont {Hosseini}}, \bibinfo {author} {\bibfnamefont
  {M.}~\bibnamefont {Zmerli}}, \bibinfo {author} {\bibfnamefont
  {R.}~\bibnamefont {Guillemin}}, \bibinfo {author} {\bibfnamefont {M.-N.}\
  \bibnamefont {Piancastelli}}, \bibinfo {author} {\bibfnamefont
  {M.}~\bibnamefont {Braune}}, \bibinfo {author} {\bibfnamefont {C.~D.}\
  \bibnamefont {Schr\"{o}ter}}, \bibinfo {author} {\bibfnamefont
  {T.}~\bibnamefont {Pfeifer}}, \bibinfo {author} {\bibfnamefont
  {F.}~\bibnamefont {Martín}}, \bibinfo {author} {\bibfnamefont
  {M.}~\bibnamefont {Simon}}, and\ \bibinfo {author} {\bibfnamefont
  {R.}~\bibnamefont {Moshammer}},\ }\bibfield  {title} {\bibinfo {title}
  {Dynamics of highly-ionized diiodomethane: Coulomb explosion, energy exchange
  and rotating fragments},\ }\href {https://doi.org/10.1088/1361-6455/adc963}
  {\bibfield  {journal} {\bibinfo  {journal} {J. Phys. B: At. Mol. Opt. Phys.}\
  }\textbf {\bibinfo {volume} {58}},\ \bibinfo {pages} {085101} (\bibinfo
  {year} {2025})}\BibitemShut {NoStop}%
\bibitem [{\citenamefont {Guo}\ \emph {et~al.}(2024)\citenamefont {Guo},
  \citenamefont {Hu}, \citenamefont {Li}, \citenamefont {Jia}, \citenamefont
  {Zhang}, \citenamefont {Cao}, \citenamefont {Xie}, \citenamefont {Cao},
  \citenamefont {Liu}, \citenamefont {Zhou}, \citenamefont {Wu}, \citenamefont
  {Wang}, and\ \citenamefont {Lu}}]{Keyu:ultrafastsci4:0073}%
  \BibitemOpen
  \bibfield  {author} {\bibinfo {author} {\bibfnamefont {K.}~\bibnamefont
  {Guo}}, \bibinfo {author} {\bibfnamefont {X.}~\bibnamefont {Hu}}, \bibinfo
  {author} {\bibfnamefont {M.}~\bibnamefont {Li}}, \bibinfo {author}
  {\bibfnamefont {C.-C.}\ \bibnamefont {Jia}}, \bibinfo {author} {\bibfnamefont
  {S.}~\bibnamefont {Zhang}}, \bibinfo {author} {\bibfnamefont
  {C.}~\bibnamefont {Cao}}, \bibinfo {author} {\bibfnamefont {W.}~\bibnamefont
  {Xie}}, \bibinfo {author} {\bibfnamefont {W.}~\bibnamefont {Cao}}, \bibinfo
  {author} {\bibfnamefont {K.}~\bibnamefont {Liu}}, \bibinfo {author}
  {\bibfnamefont {Y.}~\bibnamefont {Zhou}}, \bibinfo {author} {\bibfnamefont
  {Y.}~\bibnamefont {Wu}}, \bibinfo {author} {\bibfnamefont {J.}~\bibnamefont
  {Wang}}, and\ \bibinfo {author} {\bibfnamefont {P.}~\bibnamefont {Lu}},\
  }\bibfield  {title} {\bibinfo {title} {Probing coupled rotational and
  electronic dynamics during laser-induced molecular fragmentation},\ }\href
  {https://doi.org/10.34133/ultrafastscience.0073} {\bibfield  {journal}
  {\bibinfo  {journal} {Ultrafast Sci.}\ }\textbf {\bibinfo {volume} {4}},\
  \bibinfo {pages} {0073} (\bibinfo {year} {2024})}\BibitemShut {NoStop}%
\bibitem [{\citenamefont {Jiang}\ \emph {et~al.}(2022)\citenamefont {Jiang},
  \citenamefont {Lu}, and\ \citenamefont {Gao}}]{Jiang:JPC156:191101}%
  \BibitemOpen
  \bibfield  {author} {\bibinfo {author} {\bibfnamefont {P.}~\bibnamefont
  {Jiang}}, \bibinfo {author} {\bibfnamefont {L.}~\bibnamefont {Lu}}, and\
  \bibinfo {author} {\bibfnamefont {H.}~\bibnamefont {Gao}},\ }\bibfield
  {title} {\bibinfo {title} {Observation of rotationally dependent
  fine-structure branching ratios near the predissociation threshold {N}
  ($^2${D}$_{5/2, 3/2}$)+ {N} ($^2${D}$_{5/2, 3/2}$) of $^{14}${N}$_2$},\
  }\href {https://doi.org/https://doi.org/10.1063/5.0093426} {\bibfield
  {journal} {\bibinfo  {journal} {J. Phys. Chem.}\ }\textbf {\bibinfo {volume}
  {156}},\ \bibinfo {pages} {191101} (\bibinfo {year} {2022})}\BibitemShut
  {NoStop}%
\bibitem [{\citenamefont {Liu}\ \emph {et~al.}(2026)\citenamefont {Liu},
  \citenamefont {Jia}, \citenamefont {Ge}, \citenamefont {Li}, \citenamefont
  {Hu}, \citenamefont {Guo}, \citenamefont {Cao}, \citenamefont {Wu},
  \citenamefont {Wang}, and\ \citenamefont {Lu}}]{Liu:PRL136:053201}%
  \BibitemOpen
  \bibfield  {author} {\bibinfo {author} {\bibfnamefont {Y.}~\bibnamefont
  {Liu}}, \bibinfo {author} {\bibfnamefont {C.-C.}\ \bibnamefont {Jia}},
  \bibinfo {author} {\bibfnamefont {P.}~\bibnamefont {Ge}}, \bibinfo {author}
  {\bibfnamefont {M.}~\bibnamefont {Li}}, \bibinfo {author} {\bibfnamefont
  {X.}~\bibnamefont {Hu}}, \bibinfo {author} {\bibfnamefont {K.}~\bibnamefont
  {Guo}}, \bibinfo {author} {\bibfnamefont {W.}~\bibnamefont {Cao}}, \bibinfo
  {author} {\bibfnamefont {Y.}~\bibnamefont {Wu}}, \bibinfo {author}
  {\bibfnamefont {J.}~\bibnamefont {Wang}}, and\ \bibinfo {author}
  {\bibfnamefont {P.}~\bibnamefont {Lu}},\ }\bibfield  {title} {\bibinfo
  {title} {Radial coupling at conical intersection governs competing
  fragmentation pathways in halomethane cations},\ }\href
  {https://doi.org/10.1103/3xwt-z6bv} {\bibfield  {journal} {\bibinfo
  {journal} {Phys. Rev. Lett.}\ }\textbf {\bibinfo {volume} {136}},\ \bibinfo
  {pages} {053201} (\bibinfo {year} {2026})}\BibitemShut {NoStop}%
\bibitem [{\citenamefont {van~de Meerakker}\ \emph {et~al.}(2012)\citenamefont
  {van~de Meerakker}, \citenamefont {Bethlem}, \citenamefont {Vanhaecke}, and\
  \citenamefont {Meijer}}]{Meerakker:CR112:4828}%
  \BibitemOpen
  \bibfield  {author} {\bibinfo {author} {\bibfnamefont {S.~Y.~T.}\
  \bibnamefont {van~de Meerakker}}, \bibinfo {author} {\bibfnamefont {H.~L.}\
  \bibnamefont {Bethlem}}, \bibinfo {author} {\bibfnamefont {N.}~\bibnamefont
  {Vanhaecke}}, and\ \bibinfo {author} {\bibfnamefont {G.}~\bibnamefont
  {Meijer}},\ }\bibfield  {title} {\bibinfo {title} {Manipulation and control
  of molecular beams},\ }\href {https://doi.org/10.1021/cr200349r} {\bibfield
  {journal} {\bibinfo  {journal} {Chem. Rev.}\ }\textbf {\bibinfo {volume}
  {112}},\ \bibinfo {pages} {4828} (\bibinfo {year} {2012})}\BibitemShut
  {NoStop}%
\bibitem [{\citenamefont {Chang}\ \emph {et~al.}(2015)\citenamefont {Chang},
  \citenamefont {Horke}, \citenamefont {Trippel}, and\ \citenamefont
  {Küpper}}]{Chang:IRPC34:557}%
  \BibitemOpen
  \bibfield  {author} {\bibinfo {author} {\bibfnamefont {Y.-P.}\ \bibnamefont
  {Chang}}, \bibinfo {author} {\bibfnamefont {D.~A.}\ \bibnamefont {Horke}},
  \bibinfo {author} {\bibfnamefont {S.}~\bibnamefont {Trippel}}, and\ \bibinfo
  {author} {\bibfnamefont {J.}~\bibnamefont {Küpper}},\ }\bibfield  {title}
  {\bibinfo {title} {Spatially-controlled complex molecules and their
  applications},\ }\href {https://doi.org/10.1080/0144235X.2015.1077838}
  {\bibfield  {journal} {\bibinfo  {journal} {Int. Rev. Phys. Chem.}\ }\textbf
  {\bibinfo {volume} {34}},\ \bibinfo {pages} {557} (\bibinfo {year} {2015})},\
  \Eprint {https://arxiv.org/abs/1505.05632} {arXiv:1505.05632 [physics]}
  \BibitemShut {NoStop}%
\bibitem [{\citenamefont {Chang}\ \emph {et~al.}(2014)\citenamefont {Chang},
  \citenamefont {Filsinger}, \citenamefont {Sartakov}, and\ \citenamefont
  {K{\"u}pper}}]{Chang:CPC185:339}%
  \BibitemOpen
  \bibfield  {author} {\bibinfo {author} {\bibfnamefont {Y.-P.}\ \bibnamefont
  {Chang}}, \bibinfo {author} {\bibfnamefont {F.}~\bibnamefont {Filsinger}},
  \bibinfo {author} {\bibfnamefont {B.~G.}\ \bibnamefont {Sartakov}}, and\
  \bibinfo {author} {\bibfnamefont {J.}~\bibnamefont {K{\"u}pper}},\ }\bibfield
   {title} {\bibinfo {title} {\textsc{CMIstark}: {P}ython package for the
  {S}tark-effect calculation and symmetry classification of linear, symmetric
  and asymmetric top wavefunctions in dc electric fields},\ }\href
  {https://doi.org/10.1016/j.cpc.2013.09.001} {\bibfield  {journal} {\bibinfo
  {journal} {Comp. Phys. Comm.}\ }\textbf {\bibinfo {volume} {185}},\ \bibinfo
  {pages} {339} (\bibinfo {year} {2014})},\ \bibinfo {note} {current version
  available from \url{https://gitlab.desy.de/CMI/CMI-public/cmistark}},\
  \Eprint {https://arxiv.org/abs/1308.4076} {arXiv:1308.4076 [physics]}
  \BibitemShut {NoStop}%
\bibitem [{\citenamefont {Trippel}\ \emph {et~al.}(2018)\citenamefont
  {Trippel}, \citenamefont {Johny}, \citenamefont {Kierspel}, \citenamefont
  {Onvlee}, \citenamefont {Bieker}, \citenamefont {Ye}, \citenamefont
  {Mullins}, \citenamefont {Gumprecht}, \citenamefont {D{\l}ugo{\l}\k{e}cki},\
  and\ \citenamefont {K{\"u}pper}}]{Trippel:RSI89:096110}%
  \BibitemOpen
  \bibfield  {author} {\bibinfo {author} {\bibfnamefont {S.}~\bibnamefont
  {Trippel}}, \bibinfo {author} {\bibfnamefont {M.}~\bibnamefont {Johny}},
  \bibinfo {author} {\bibfnamefont {T.}~\bibnamefont {Kierspel}}, \bibinfo
  {author} {\bibfnamefont {J.}~\bibnamefont {Onvlee}}, \bibinfo {author}
  {\bibfnamefont {H.}~\bibnamefont {Bieker}}, \bibinfo {author} {\bibfnamefont
  {H.}~\bibnamefont {Ye}}, \bibinfo {author} {\bibfnamefont {T.}~\bibnamefont
  {Mullins}}, \bibinfo {author} {\bibfnamefont {L.}~\bibnamefont {Gumprecht}},
  \bibinfo {author} {\bibfnamefont {K.}~\bibnamefont {D{\l}ugo{\l}\k{e}cki}},\
  and\ \bibinfo {author} {\bibfnamefont {J.}~\bibnamefont {K{\"u}pper}},\
  }\bibfield  {title} {\bibinfo {title} {Note: Knife edge skimming for improved
  separation of molecular species by the deflector},\ }\href
  {https://doi.org/10.1063/1.5026145} {\bibfield  {journal} {\bibinfo
  {journal} {Rev. Sci. Instrum.}\ }\textbf {\bibinfo {volume} {89}},\ \bibinfo
  {pages} {096110} (\bibinfo {year} {2018})},\ \Eprint
  {https://arxiv.org/abs/1802.04053} {arXiv:1802.04053 [physics]}\BibitemShut
  {NoStop}%
\bibitem [{\citenamefont {Filsinger}\ \emph {et~al.}(2008)\citenamefont
  {Filsinger}, \citenamefont {Erlekam}, \citenamefont {von Helden},
  \citenamefont {K{\"u}pper}, and\ \citenamefont
  {Meijer}}]{Filsinger:PRL100:133003}%
  \BibitemOpen
  \bibfield  {author} {\bibinfo {author} {\bibfnamefont {F.}~\bibnamefont
  {Filsinger}}, \bibinfo {author} {\bibfnamefont {U.}~\bibnamefont {Erlekam}},
  \bibinfo {author} {\bibfnamefont {G.}~\bibnamefont {von Helden}}, \bibinfo
  {author} {\bibfnamefont {J.}~\bibnamefont {K{\"u}pper}}, and\ \bibinfo
  {author} {\bibfnamefont {G.}~\bibnamefont {Meijer}},\ }\bibfield  {title}
  {\bibinfo {title} {Selector for structural isomers of neutral molecules},\
  }\href {https://doi.org/10.1103/PhysRevLett.100.133003} {\bibfield  {journal}
  {\bibinfo  {journal} {Phys. Rev. Lett.}\ }\textbf {\bibinfo {volume} {100}},\
  \bibinfo {pages} {133003} (\bibinfo {year} {2008})},\ \Eprint
  {https://arxiv.org/abs/0802.2795} {arXiv:0802.2795 [physics]}\BibitemShut
  {NoStop}%
\bibitem [{\citenamefont {He}\ \emph {et~al.}(2024)\citenamefont {He},
  \citenamefont {Johny}, \citenamefont {Kierspel}, \citenamefont
  {Długołęcki}, \citenamefont {Bari}, \citenamefont {Boll}, \citenamefont
  {Bromberger}, \citenamefont {Coreno}, \citenamefont {Fanis}, \citenamefont
  {Fraia}, \citenamefont {Erk}, \citenamefont {Gisselbrecht}, \citenamefont
  {Grychtol}, \citenamefont {Eng-Johnsson}, \citenamefont {Mazza},
  \citenamefont {Onvlee}, \citenamefont {Ovcharenko}, \citenamefont {Petrovic},
  \citenamefont {Rennhack}, \citenamefont {Rivas}, \citenamefont {Rudenko},
  \citenamefont {Rühl}, \citenamefont {Schwob}, \citenamefont {Simon},
  \citenamefont {Trinter}, \citenamefont {Usenko}, \citenamefont {Wiese},
  \citenamefont {Meyer}, \citenamefont {Trippel}, and\ \citenamefont
  {Küpper}}]{He:RSI95:113301}%
  \BibitemOpen
  \bibfield  {author} {\bibinfo {author} {\bibfnamefont {L.}~\bibnamefont
  {He}}, \bibinfo {author} {\bibfnamefont {M.}~\bibnamefont {Johny}}, \bibinfo
  {author} {\bibfnamefont {T.}~\bibnamefont {Kierspel}}, \bibinfo {author}
  {\bibfnamefont {K.}~\bibnamefont {Długołęcki}}, \bibinfo {author}
  {\bibfnamefont {S.}~\bibnamefont {Bari}}, \bibinfo {author} {\bibfnamefont
  {R.}~\bibnamefont {Boll}}, \bibinfo {author} {\bibfnamefont {H.}~\bibnamefont
  {Bromberger}}, \bibinfo {author} {\bibfnamefont {M.}~\bibnamefont {Coreno}},
  \bibinfo {author} {\bibfnamefont {A.~D.}\ \bibnamefont {Fanis}}, \bibinfo
  {author} {\bibfnamefont {M.~D.}\ \bibnamefont {Fraia}}, \bibinfo {author}
  {\bibfnamefont {B.}~\bibnamefont {Erk}}, \bibinfo {author} {\bibfnamefont
  {M.}~\bibnamefont {Gisselbrecht}}, \bibinfo {author} {\bibfnamefont
  {P.}~\bibnamefont {Grychtol}}, \bibinfo {author} {\bibfnamefont
  {P.}~\bibnamefont {Eng-Johnsson}}, \bibinfo {author} {\bibfnamefont
  {T.}~\bibnamefont {Mazza}}, \bibinfo {author} {\bibfnamefont
  {J.}~\bibnamefont {Onvlee}}, \bibinfo {author} {\bibfnamefont
  {Y.}~\bibnamefont {Ovcharenko}}, \bibinfo {author} {\bibfnamefont
  {J.}~\bibnamefont {Petrovic}}, \bibinfo {author} {\bibfnamefont
  {N.}~\bibnamefont {Rennhack}}, \bibinfo {author} {\bibfnamefont {D.~E.}\
  \bibnamefont {Rivas}}, \bibinfo {author} {\bibfnamefont {A.}~\bibnamefont
  {Rudenko}}, \bibinfo {author} {\bibfnamefont {E.}~\bibnamefont {Rühl}},
  \bibinfo {author} {\bibfnamefont {L.}~\bibnamefont {Schwob}}, \bibinfo
  {author} {\bibfnamefont {M.}~\bibnamefont {Simon}}, \bibinfo {author}
  {\bibfnamefont {F.}~\bibnamefont {Trinter}}, \bibinfo {author} {\bibfnamefont
  {S.}~\bibnamefont {Usenko}}, \bibinfo {author} {\bibfnamefont
  {J.}~\bibnamefont {Wiese}}, \bibinfo {author} {\bibfnamefont
  {M.}~\bibnamefont {Meyer}}, \bibinfo {author} {\bibfnamefont
  {S.}~\bibnamefont {Trippel}}, and\ \bibinfo {author} {\bibfnamefont
  {J.}~\bibnamefont {Küpper}},\ }\bibfield  {title} {\bibinfo {title}
  {Controlled molecule injector for cold, dense, and pure molecular beams at
  the european x-ray free-electron laser},\ }\href
  {https://doi.org/10.1063/5.0219086} {\bibfield  {journal} {\bibinfo
  {journal} {Rev. Sci. Instrum.}\ }\textbf {\bibinfo {volume} {95}},\ \bibinfo
  {pages} {113301} (\bibinfo {year} {2024})},\ \Eprint
  {https://arxiv.org/abs/2405.06344} {arXiv:2405.06344 [physics]}\BibitemShut
  {NoStop}%
\bibitem [{\citenamefont {Chang}\ \emph {et~al.}(2013)\citenamefont {Chang},
  \citenamefont {D{\l}ugo\l\k{e}cki}, \citenamefont {K{\"u}pper}, \citenamefont
  {R{\"o}sch}, \citenamefont {Wild}, and\ \citenamefont
  {Willitsch}}]{Chang:Science342:98}%
  \BibitemOpen
  \bibfield  {author} {\bibinfo {author} {\bibfnamefont {Y.-P.}\ \bibnamefont
  {Chang}}, \bibinfo {author} {\bibfnamefont {K.}~\bibnamefont
  {D{\l}ugo\l\k{e}cki}}, \bibinfo {author} {\bibfnamefont {J.}~\bibnamefont
  {K{\"u}pper}}, \bibinfo {author} {\bibfnamefont {D.}~\bibnamefont
  {R{\"o}sch}}, \bibinfo {author} {\bibfnamefont {D.}~\bibnamefont {Wild}},\
  and\ \bibinfo {author} {\bibfnamefont {S.}~\bibnamefont {Willitsch}},\
  }\bibfield  {title} {\bibinfo {title} {Specific chemical reactivities of
  spatially separated 3-aminophenol conformers with cold {Ca$^+$} ions},\
  }\href {https://doi.org/10.1126/science.1242271} {\bibfield  {journal}
  {\bibinfo  {journal} {Science}\ }\textbf {\bibinfo {volume} {342}},\ \bibinfo
  {pages} {98} (\bibinfo {year} {2013})},\ \Eprint
  {https://arxiv.org/abs/1308.6538} {arXiv:1308.6538 [physics]}\BibitemShut
  {NoStop}%
\bibitem [{\citenamefont {Vinkl\'{a}rek}\ \emph {et~al.}(2024)\citenamefont
  {Vinkl\'{a}rek}, \citenamefont {Bromberger}, \citenamefont {Vadassery},
  \citenamefont {Jin}, \citenamefont {Küpper}, and\ \citenamefont
  {Trippel}}]{Vinklarek:JPCA128:1593}%
  \BibitemOpen
  \bibfield  {author} {\bibinfo {author} {\bibfnamefont {I.~S.}\ \bibnamefont
  {Vinkl\'{a}rek}}, \bibinfo {author} {\bibfnamefont {H.}~\bibnamefont
  {Bromberger}}, \bibinfo {author} {\bibfnamefont {N.}~\bibnamefont
  {Vadassery}}, \bibinfo {author} {\bibfnamefont {W.}~\bibnamefont {Jin}},
  \bibinfo {author} {\bibfnamefont {J.}~\bibnamefont {Küpper}}, and\ \bibinfo
  {author} {\bibfnamefont {S.}~\bibnamefont {Trippel}},\ }\bibfield  {title}
  {\bibinfo {title} {Reaction pathways of water dimer following single
  ionization},\ }\href {https://doi.org/10.1021/acs.jpca.3c07958} {\bibfield
  {journal} {\bibinfo  {journal} {J. Phys. Chem. A}\ }\textbf {\bibinfo
  {volume} {128}},\ \bibinfo {pages} {1593} (\bibinfo {year} {2024})},\ \Eprint
  {https://arxiv.org/abs/2308.08006} {arXiv:2308.08006 [physics]}\BibitemShut
  {NoStop}%
\bibitem [{\citenamefont {Johny}\ \emph {et~al.}(2024)\citenamefont {Johny},
  \citenamefont {Schouder}, \citenamefont {Al-Refaie}, \citenamefont {He},
  \citenamefont {Wiese}, \citenamefont {Stapelfeldt}, \citenamefont {Trippel},\
  and\ \citenamefont {Küpper}}]{Johny:PCCP26:13118}%
  \BibitemOpen
  \bibfield  {author} {\bibinfo {author} {\bibfnamefont {M.}~\bibnamefont
  {Johny}}, \bibinfo {author} {\bibfnamefont {C.~A.}\ \bibnamefont {Schouder}},
  \bibinfo {author} {\bibfnamefont {A.}~\bibnamefont {Al-Refaie}}, \bibinfo
  {author} {\bibfnamefont {L.}~\bibnamefont {He}}, \bibinfo {author}
  {\bibfnamefont {J.}~\bibnamefont {Wiese}}, \bibinfo {author} {\bibfnamefont
  {H.}~\bibnamefont {Stapelfeldt}}, \bibinfo {author} {\bibfnamefont
  {S.}~\bibnamefont {Trippel}}, and\ \bibinfo {author} {\bibfnamefont
  {J.}~\bibnamefont {Küpper}},\ }\bibfield  {title} {\bibinfo {title} {Water
  is a radiation protection agent for ionised pyrrole},\ }\href
  {https://doi.org/10.1039/D3CP03471B} {\bibfield  {journal} {\bibinfo
  {journal} {Phys. Chem. Chem. Phys.}\ }\textbf {\bibinfo {volume} {26}},\
  \bibinfo {pages} {13118} (\bibinfo {year} {2024})},\ \Eprint
  {https://arxiv.org/abs/2010.00453} {arXiv:2010.00453 [physics]}\BibitemShut
  {NoStop}%
\bibitem [{\citenamefont {Kilaj}\ \emph {et~al.}(2021)\citenamefont {Kilaj},
  \citenamefont {Wang}, \citenamefont {Straň{\'{a}}k}, \citenamefont
  {Schwilk}, \citenamefont {Rivero}, \citenamefont {Xu}, \citenamefont {von
  Lilienfeld}, \citenamefont {K{\"{u}}pper}, and\ \citenamefont
  {Willitsch}}]{Kilaj:NatComm12:6047}%
  \BibitemOpen
  \bibfield  {author} {\bibinfo {author} {\bibfnamefont {A.}~\bibnamefont
  {Kilaj}}, \bibinfo {author} {\bibfnamefont {J.}~\bibnamefont {Wang}},
  \bibinfo {author} {\bibfnamefont {P.}~\bibnamefont {Straň{\'{a}}k}},
  \bibinfo {author} {\bibfnamefont {M.}~\bibnamefont {Schwilk}}, \bibinfo
  {author} {\bibfnamefont {U.}~\bibnamefont {Rivero}}, \bibinfo {author}
  {\bibfnamefont {L.}~\bibnamefont {Xu}}, \bibinfo {author} {\bibfnamefont
  {O.~A.}\ \bibnamefont {von Lilienfeld}}, \bibinfo {author} {\bibfnamefont
  {J.}~\bibnamefont {K{\"{u}}pper}}, and\ \bibinfo {author} {\bibfnamefont
  {S.}~\bibnamefont {Willitsch}},\ }\bibfield  {title} {\bibinfo {title}
  {{Conformer-specific polar cycloaddition of dibromobutadiene with trapped
  propene ions}},\ }\href {https://doi.org/10.1038/s41467-021-26309-5}
  {\bibfield  {journal} {\bibinfo  {journal} {Nat. Commun.}\ }\textbf {\bibinfo
  {volume} {12}},\ \bibinfo {pages} {6047} (\bibinfo {year} {2021})},\ \Eprint
  {https://arxiv.org/abs/2107.13858} {2107.13858}\BibitemShut {NoStop}%
\bibitem [{\citenamefont {Kilaj}\ \emph {et~al.}(2018)\citenamefont {Kilaj},
  \citenamefont {Gao}, \citenamefont {R\"osch}, \citenamefont {Rivero},
  \citenamefont {K\"upper}, and\ \citenamefont
  {Willitsch}}]{Kilaj:NatComm9:2096}%
  \BibitemOpen
  \bibfield  {author} {\bibinfo {author} {\bibfnamefont {A.}~\bibnamefont
  {Kilaj}}, \bibinfo {author} {\bibfnamefont {H.}~\bibnamefont {Gao}}, \bibinfo
  {author} {\bibfnamefont {D.}~\bibnamefont {R\"osch}}, \bibinfo {author}
  {\bibfnamefont {U.}~\bibnamefont {Rivero}}, \bibinfo {author} {\bibfnamefont
  {J.}~\bibnamefont {K\"upper}}, and\ \bibinfo {author} {\bibfnamefont
  {S.}~\bibnamefont {Willitsch}},\ }\bibfield  {title} {\bibinfo {title}
  {Observation of different reactivities of para- and ortho-water towards
  trapped diazenylium ions},\ }\href
  {https://doi.org/10.1038/s41467-018-04483-3} {\bibfield  {journal} {\bibinfo
  {journal} {Nat. Commun.}\ }\textbf {\bibinfo {volume} {9}},\ \bibinfo {pages}
  {2096} (\bibinfo {year} {2018})}\BibitemShut {NoStop}%
\bibitem [{\citenamefont {Jin}\ \emph {et~al.}(2025)\citenamefont {Jin},
  \citenamefont {Bromberger}, \citenamefont {He}, \citenamefont {Johny},
  \citenamefont {Vinkl{\'a}rek}, \citenamefont {D{\l}ugo{\l}{\k{e}}cki},
  \citenamefont {Samartsev}, \citenamefont {Calegari}, \citenamefont
  {Trippel}, and\ \citenamefont {K{\"u}pper}}]{Jin:RSI96:023305}%
  \BibitemOpen
  \bibfield  {author} {\bibinfo {author} {\bibfnamefont {W.}~\bibnamefont
  {Jin}}, \bibinfo {author} {\bibfnamefont {H.}~\bibnamefont {Bromberger}},
  \bibinfo {author} {\bibfnamefont {L.}~\bibnamefont {He}}, \bibinfo {author}
  {\bibfnamefont {M.}~\bibnamefont {Johny}}, \bibinfo {author} {\bibfnamefont
  {I.~S.}\ \bibnamefont {Vinkl{\'a}rek}}, \bibinfo {author} {\bibfnamefont
  {K.}~\bibnamefont {D{\l}ugo{\l}{\k{e}}cki}}, \bibinfo {author} {\bibfnamefont
  {A.}~\bibnamefont {Samartsev}}, \bibinfo {author} {\bibfnamefont
  {F.}~\bibnamefont {Calegari}}, \bibinfo {author} {\bibfnamefont
  {S.}~\bibnamefont {Trippel}}, and\ \bibinfo {author} {\bibfnamefont
  {J.}~\bibnamefont {K{\"u}pper}},\ }\bibfield  {title} {\bibinfo {title} {A
  versatile and transportable endstation for controlled molecule experiments},\
  }\href {https://doi.org/10.1063/5.0228913} {\bibfield  {journal} {\bibinfo
  {journal} {Rev. Sci. Instrum.}\ }\textbf {\bibinfo {volume} {96}},\ \bibinfo
  {pages} {023305} (\bibinfo {year} {2025})}\BibitemShut {NoStop}%
\bibitem [{\citenamefont {Even}(2015)}]{Even:EPJTI2:17}%
  \BibitemOpen
  \bibfield  {author} {\bibinfo {author} {\bibfnamefont {U.}~\bibnamefont
  {Even}},\ }\bibfield  {title} {\bibinfo {title} {The {E}ven-{L}avie valve as
  a source for high intensity supersonic beam},\ }\href
  {https://doi.org/10.1140/epjti/s40485-015-0027-5} {\bibfield  {journal}
  {\bibinfo  {journal} {Eur. Phys. J. Techn. Instrumen.}\ }\textbf {\bibinfo
  {volume} {2}},\ \bibinfo {pages} {17} (\bibinfo {year} {2015})}\BibitemShut
  {NoStop}%
\bibitem [{\citenamefont {Bromberger}\ \emph {et~al.}(2022)\citenamefont
  {Bromberger}, \citenamefont {Passow}, \citenamefont {Pennicard},
  \citenamefont {Boll}, \citenamefont {Correa}, \citenamefont {He},
  \citenamefont {Johny}, \citenamefont {Papadopoulou}, \citenamefont
  {Tul-Noor}, \citenamefont {Wiese}, \citenamefont {Trippel}, \citenamefont
  {Erk}, and\ \citenamefont {Küpper}}]{Bromberger:JPB55:144001}%
  \BibitemOpen
  \bibfield  {author} {\bibinfo {author} {\bibfnamefont {H.}~\bibnamefont
  {Bromberger}}, \bibinfo {author} {\bibfnamefont {C.}~\bibnamefont {Passow}},
  \bibinfo {author} {\bibfnamefont {D.}~\bibnamefont {Pennicard}}, \bibinfo
  {author} {\bibfnamefont {R.}~\bibnamefont {Boll}}, \bibinfo {author}
  {\bibfnamefont {J.}~\bibnamefont {Correa}}, \bibinfo {author} {\bibfnamefont
  {L.}~\bibnamefont {He}}, \bibinfo {author} {\bibfnamefont {M.}~\bibnamefont
  {Johny}}, \bibinfo {author} {\bibfnamefont {C.}~\bibnamefont {Papadopoulou}},
  \bibinfo {author} {\bibfnamefont {A.}~\bibnamefont {Tul-Noor}}, \bibinfo
  {author} {\bibfnamefont {J.}~\bibnamefont {Wiese}}, \bibinfo {author}
  {\bibfnamefont {S.}~\bibnamefont {Trippel}}, \bibinfo {author} {\bibfnamefont
  {B.}~\bibnamefont {Erk}}, and\ \bibinfo {author} {\bibfnamefont
  {J.}~\bibnamefont {Küpper}},\ }\bibfield  {title} {\bibinfo {title}
  {{Shot-by-shot 250~kHz 3D ion and MHz photoelectron imaging using
  Timepix3}},\ }\href {https://doi.org/10.1088/1361-6455/ac6b6b} {\bibfield
  {journal} {\bibinfo  {journal} {J. Phys. B}\ }\textbf {\bibinfo {volume}
  {55}},\ \bibinfo {pages} {144001} (\bibinfo {year} {2022})},\ \Eprint
  {https://arxiv.org/abs/2111.14407} {arXiv:2111.14407 [physics]}\BibitemShut
  {NoStop}%
\bibitem [{\citenamefont {L'Huillier}\ \emph {et~al.}(1983)\citenamefont
  {L'Huillier}, \citenamefont {Lompre}, \citenamefont {Mainfray}, and\
  \citenamefont {Manus}}]{LHuillier:PRA27:2503}%
  \BibitemOpen
  \bibfield  {author} {\bibinfo {author} {\bibfnamefont {A.}~\bibnamefont
  {L'Huillier}}, \bibinfo {author} {\bibfnamefont {L.~A.}\ \bibnamefont
  {Lompre}}, \bibinfo {author} {\bibfnamefont {G.}~\bibnamefont {Mainfray}},\
  and\ \bibinfo {author} {\bibfnamefont {C.}~\bibnamefont {Manus}},\ }\bibfield
   {title} {\bibinfo {title} {Multiply charged ions induced by multiphoton
  absorption in rare-gases at 0.53~$\mathrm{\mu m}$},\ }\href
  {https://doi.org/10.1103/PhysRevA.27.2503} {\bibfield  {journal} {\bibinfo
  {journal} {Phys. Rev. A}\ }\textbf {\bibinfo {volume} {27}},\ \bibinfo
  {pages} {2503} (\bibinfo {year} {1983})}\BibitemShut {NoStop}%
\bibitem [{\citenamefont {Wiese}\ \emph {et~al.}(2019)\citenamefont {Wiese},
  \citenamefont {Olivieri}, \citenamefont {Trabattoni}, \citenamefont
  {Trippel}, and\ \citenamefont {Küpper}}]{Wiese:NJP21:083011}%
  \BibitemOpen
  \bibfield  {author} {\bibinfo {author} {\bibfnamefont {J.}~\bibnamefont
  {Wiese}}, \bibinfo {author} {\bibfnamefont {J.-F.}\ \bibnamefont {Olivieri}},
  \bibinfo {author} {\bibfnamefont {A.}~\bibnamefont {Trabattoni}}, \bibinfo
  {author} {\bibfnamefont {S.}~\bibnamefont {Trippel}}, and\ \bibinfo {author}
  {\bibfnamefont {J.}~\bibnamefont {Küpper}},\ }\bibfield  {title} {\bibinfo
  {title} {Strong-field photoelectron momentum imaging of {OCS} at finely
  resolved incident intensities},\ }\href
  {https://doi.org/10.1088/1367-2630/ab34e8} {\bibfield  {journal} {\bibinfo
  {journal} {New J. Phys.}\ }\textbf {\bibinfo {volume} {21}},\ \bibinfo
  {pages} {083011} (\bibinfo {year} {2019})},\ \Eprint
  {https://arxiv.org/abs/1904.07519} {arXiv:1904.07519 [physics]}\BibitemShut
  {NoStop}%
\bibitem [{\citenamefont {Vinkl{\'a}rek}\ \emph {et~al.}(2026)\citenamefont
  {Vinkl{\'a}rek}, \citenamefont {Vadassery}, \citenamefont {Pradhan},
  \citenamefont {Bromberger}, \citenamefont {Küpper}, and\ \citenamefont
  {Trippel}}]{Vinklarek:SlowFragmentationDynamics:inprep}%
  \BibitemOpen
  \bibfield  {author} {\bibinfo {author} {\bibfnamefont {I.~S.}\ \bibnamefont
  {Vinkl{\'a}rek}}, \bibinfo {author} {\bibfnamefont {N.}~\bibnamefont
  {Vadassery}}, \bibinfo {author} {\bibfnamefont {A.}~\bibnamefont {Pradhan}},
  \bibinfo {author} {\bibfnamefont {H.}~\bibnamefont {Bromberger}}, \bibinfo
  {author} {\bibfnamefont {J.}~\bibnamefont {Küpper}}, and\ \bibinfo {author}
  {\bibfnamefont {S.}~\bibnamefont {Trippel}},\ }\href@noop {} {\bibinfo
  {title} {Direct imaging of slow fragmentation dynamics demonstrated on
  {CF}$_2${I}$_2$ evaporation}} (\bibinfo {year} {2026}),\ \bibinfo {note} {in
  preparation}\BibitemShut {NoStop}%
\bibitem [{\citenamefont {Filsinger}\ \emph {et~al.}(2009)\citenamefont
  {Filsinger}, \citenamefont {K{\"u}pper}, \citenamefont {Meijer},
  \citenamefont {Holmegaard}, \citenamefont {Nielsen}, \citenamefont {Nevo},
  \citenamefont {Hansen}, and\ \citenamefont
  {Stapelfeldt}}]{Filsinger:JCP131:064309}%
  \BibitemOpen
  \bibfield  {author} {\bibinfo {author} {\bibfnamefont {F.}~\bibnamefont
  {Filsinger}}, \bibinfo {author} {\bibfnamefont {J.}~\bibnamefont
  {K{\"u}pper}}, \bibinfo {author} {\bibfnamefont {G.}~\bibnamefont {Meijer}},
  \bibinfo {author} {\bibfnamefont {L.}~\bibnamefont {Holmegaard}}, \bibinfo
  {author} {\bibfnamefont {J.~H.}\ \bibnamefont {Nielsen}}, \bibinfo {author}
  {\bibfnamefont {I.}~\bibnamefont {Nevo}}, \bibinfo {author} {\bibfnamefont
  {J.~L.}\ \bibnamefont {Hansen}}, and\ \bibinfo {author} {\bibfnamefont
  {H.}~\bibnamefont {Stapelfeldt}},\ }\bibfield  {title} {\bibinfo {title}
  {Quantum-state selection, alignment, and orientation of large molecules using
  static electric and laser fields},\ }\href
  {https://doi.org/10.1063/1.3194287} {\bibfield  {journal} {\bibinfo
  {journal} {J. Chem. Phys.}\ }\textbf {\bibinfo {volume} {131}},\ \bibinfo
  {pages} {064309} (\bibinfo {year} {2009})},\ \Eprint
  {https://arxiv.org/abs/0903.5413} {arXiv:0903.5413 [physics]}\BibitemShut
  {NoStop}%
\bibitem [{\citenamefont {van Veldhoven}\ \emph {et~al.}(2002)\citenamefont
  {van Veldhoven}, \citenamefont {Jongma}, \citenamefont {Sartakov},
  \citenamefont {Bongers}, and\ \citenamefont
  {Meij\-er}}]{Veldhoven:PRA66:032501}%
  \BibitemOpen
  \bibfield  {author} {\bibinfo {author} {\bibfnamefont {J.}~\bibnamefont {van
  Veldhoven}}, \bibinfo {author} {\bibfnamefont {R.~T.}\ \bibnamefont
  {Jongma}}, \bibinfo {author} {\bibfnamefont {B.}~\bibnamefont {Sartakov}},
  \bibinfo {author} {\bibfnamefont {W.~A.}\ \bibnamefont {Bongers}}, and\
  \bibinfo {author} {\bibfnamefont {G.}~\bibnamefont {Meij\-er}},\ }\bibfield
  {title} {\bibinfo {title} {Hyperfine structure of {ND}$_3$},\ }\href@noop {}
  {\bibfield  {journal} {\bibinfo  {journal} {Phys. Rev. A}\ }\textbf {\bibinfo
  {volume} {66}},\ \bibinfo {pages} {32501} (\bibinfo {year}
  {2002})}\BibitemShut {NoStop}%
\bibitem [{\citenamefont {Wang}\ \emph {et~al.}(2006)\citenamefont {Wang},
  \citenamefont {Luo}, \citenamefont {Wang}, \citenamefont {Hu}, \citenamefont
  {Wang}, \citenamefont {Zhao}, and\ \citenamefont
  {Zhang}}]{Wang:IJQC106:1138}%
  \BibitemOpen
  \bibfield  {author} {\bibinfo {author} {\bibfnamefont {H.}~\bibnamefont
  {Wang}}, \bibinfo {author} {\bibfnamefont {S.-Z.}\ \bibnamefont {Luo}},
  \bibinfo {author} {\bibfnamefont {Y.}~\bibnamefont {Wang}}, \bibinfo {author}
  {\bibfnamefont {M.-L.}\ \bibnamefont {Hu}}, \bibinfo {author} {\bibfnamefont
  {Q.-Y.}\ \bibnamefont {Wang}}, \bibinfo {author} {\bibfnamefont {S.-W.}\
  \bibnamefont {Zhao}}, and\ \bibinfo {author} {\bibfnamefont {J.-Y.}\
  \bibnamefont {Zhang}},\ }\bibfield  {title} {\bibinfo {title} {Field-assisted
  dissociative ionization of {CH$_2$I$_2$} induced by femtosecond laser
  field},\ }\href {https://doi.org/10.1002/qua.20874} {\bibfield  {journal}
  {\bibinfo  {journal} {Int. J. Quantum Chem.}\ }\textbf {\bibinfo {volume}
  {106}},\ \bibinfo {pages} {1138} (\bibinfo {year} {2006})}\BibitemShut
  {NoStop}%
\bibitem [{\citenamefont {Zhang}\ \emph {et~al.}(2010)\citenamefont {Zhang},
  \citenamefont {Zhang}, \citenamefont {Liu}, \citenamefont {Xu}, \citenamefont
  {Jin}, and\ \citenamefont {Ding}}]{Zhang:JPB43:025102}%
  \BibitemOpen
  \bibfield  {author} {\bibinfo {author} {\bibfnamefont {X.}~\bibnamefont
  {Zhang}}, \bibinfo {author} {\bibfnamefont {D.}~\bibnamefont {Zhang}},
  \bibinfo {author} {\bibfnamefont {H.}~\bibnamefont {Liu}}, \bibinfo {author}
  {\bibfnamefont {H.}~\bibnamefont {Xu}}, \bibinfo {author} {\bibfnamefont
  {M.}~\bibnamefont {Jin}}, and\ \bibinfo {author} {\bibfnamefont
  {D.}~\bibnamefont {Ding}},\ }\bibfield  {title} {\bibinfo {title} {Angular
  distributions of fragment ions in dissociative ionization of {CH}$_2${I}$_2$
  molecules in intense laser fields},\ }\href
  {https://doi.org/10.1088/0953-4075/43/2/025102} {\bibfield  {journal}
  {\bibinfo  {journal} {J. Phys. B}\ }\textbf {\bibinfo {volume} {43}},\
  \bibinfo {pages} {025102} (\bibinfo {year} {2010})}\BibitemShut {NoStop}%
\bibitem [{\citenamefont {Hankin}\ \emph {et~al.}(2000)\citenamefont {Hankin},
  \citenamefont {Villeneuve}, \citenamefont {Corkum}, and\ \citenamefont
  {Rayner}}]{Hankin:PRL84:5082}%
  \BibitemOpen
  \bibfield  {author} {\bibinfo {author} {\bibfnamefont {S.~M.}\ \bibnamefont
  {Hankin}}, \bibinfo {author} {\bibfnamefont {D.~M.}\ \bibnamefont
  {Villeneuve}}, \bibinfo {author} {\bibfnamefont {P.~B.}\ \bibnamefont
  {Corkum}}, and\ \bibinfo {author} {\bibfnamefont {D.~M.}\ \bibnamefont
  {Rayner}},\ }\bibfield  {title} {\bibinfo {title} {Nonlinear ionization of
  organic molecules in high intensity laser fields},\ }\href
  {https://doi.org/https://doi.org/10.1103/PhysRevLett.84.5082} {\bibfield
  {journal} {\bibinfo  {journal} {Phys. Rev. Lett.}\ }\textbf {\bibinfo
  {volume} {84}},\ \bibinfo {pages} {5082} (\bibinfo {year}
  {2000})}\BibitemShut {NoStop}%
\bibitem [{\citenamefont {Hankin}\ \emph {et~al.}(2001)\citenamefont {Hankin},
  \citenamefont {Villeneuve}, \citenamefont {Corkum}, and\ \citenamefont
  {Rayner}}]{Hankin:PRA64:013405}%
  \BibitemOpen
  \bibfield  {author} {\bibinfo {author} {\bibfnamefont {S.~M.}\ \bibnamefont
  {Hankin}}, \bibinfo {author} {\bibfnamefont {D.~M.}\ \bibnamefont
  {Villeneuve}}, \bibinfo {author} {\bibfnamefont {P.~B.}\ \bibnamefont
  {Corkum}}, and\ \bibinfo {author} {\bibfnamefont {D.~M.}\ \bibnamefont
  {Rayner}},\ }\bibfield  {title} {\bibinfo {title} {Intense-field laser
  ionization rates in atoms and molecules},\ }\href
  {https://doi.org/10.1103/PhysRevA.64.013405} {\bibfield  {journal} {\bibinfo
  {journal} {Phys. Rev. A}\ }\textbf {\bibinfo {volume} {64}},\ \bibinfo
  {pages} {013405} (\bibinfo {year} {2001})}\BibitemShut {NoStop}%
\bibitem [{\citenamefont {Leibscher}\ \emph {et~al.}(2003)\citenamefont
  {Leibscher}, \citenamefont {Averbukh}, and\ \citenamefont
  {Rabitz}}]{Leibscher:PRL90:213001}%
  \BibitemOpen
  \bibfield  {author} {\bibinfo {author} {\bibfnamefont {M.}~\bibnamefont
  {Leibscher}}, \bibinfo {author} {\bibfnamefont {I.}~\bibnamefont
  {Averbukh}}, and\ \bibinfo {author} {\bibfnamefont {H.}~\bibnamefont
  {Rabitz}},\ }\bibfield  {title} {\bibinfo {title} {Molecular alignment by
  trains of short laser pulses},\ }\href
  {https://doi.org/10.1103/PhysRevLett.90.213001} {\bibfield  {journal}
  {\bibinfo  {journal} {Phys. Rev. Lett.}\ }\textbf {\bibinfo {volume} {90}},\
  \bibinfo {pages} {213001} (\bibinfo {year} {2003})}\BibitemShut {NoStop}%
\bibitem [{\citenamefont {Karamatskos}\ \emph {et~al.}(2019)\citenamefont
  {Karamatskos}, \citenamefont {Raabe}, \citenamefont {Mullins}, \citenamefont
  {Trabattoni}, \citenamefont {Stammer}, \citenamefont {Goldsztejn},
  \citenamefont {Johansen}, \citenamefont {D{\l}ugo{\l}\k{e}cki}, \citenamefont
  {Stapelfeldt}, \citenamefont {Vrakking}, \citenamefont {Trippel},
  \citenamefont {Rouzée}, and\ \citenamefont
  {Küpper}}]{Karamatskos:NatComm10:3364}%
  \BibitemOpen
  \bibfield  {author} {\bibinfo {author} {\bibfnamefont {E.~T.}\ \bibnamefont
  {Karamatskos}}, \bibinfo {author} {\bibfnamefont {S.}~\bibnamefont {Raabe}},
  \bibinfo {author} {\bibfnamefont {T.}~\bibnamefont {Mullins}}, \bibinfo
  {author} {\bibfnamefont {A.}~\bibnamefont {Trabattoni}}, \bibinfo {author}
  {\bibfnamefont {P.}~\bibnamefont {Stammer}}, \bibinfo {author} {\bibfnamefont
  {G.}~\bibnamefont {Goldsztejn}}, \bibinfo {author} {\bibfnamefont {R.~R.}\
  \bibnamefont {Johansen}}, \bibinfo {author} {\bibfnamefont {K.}~\bibnamefont
  {D{\l}ugo{\l}\k{e}cki}}, \bibinfo {author} {\bibfnamefont {H.}~\bibnamefont
  {Stapelfeldt}}, \bibinfo {author} {\bibfnamefont {M.~J.~J.}\ \bibnamefont
  {Vrakking}}, \bibinfo {author} {\bibfnamefont {S.}~\bibnamefont {Trippel}},
  \bibinfo {author} {\bibfnamefont {A.}~\bibnamefont {Rouzée}}, and\ \bibinfo
  {author} {\bibfnamefont {J.}~\bibnamefont {Küpper}},\ }\bibfield  {title}
  {\bibinfo {title} {Molecular movie of ultrafast coherent rotational dynamics
  of {OCS}},\ }\href {https://doi.org/10.1038/s41467-019-11122-y} {\bibfield
  {journal} {\bibinfo  {journal} {Nat. Commun.}\ }\textbf {\bibinfo {volume}
  {10}},\ \bibinfo {pages} {3364} (\bibinfo {year} {2019})},\ \Eprint
  {https://arxiv.org/abs/1807.01034} {arXiv:1807.01034 [physics]}\BibitemShut
  {NoStop}%
\bibitem [{\citenamefont {Brown} and\ \citenamefont
  {Watson}(1977)}]{Brown:JMS65:65}%
  \BibitemOpen
  \bibfield  {author} {\bibinfo {author} {\bibfnamefont {J.~M.}\ \bibnamefont
  {Brown}} and\ \bibinfo {author} {\bibfnamefont {J.~K.}\ \bibnamefont
  {Watson}},\ }\bibfield  {title} {\bibinfo {title} {Spin-orbit and
  spin-rotation coupling in doublet states of diatomic molecules},\ }\href
  {https://doi.org/https://doi.org/10.1016/0022-2852(77)90358-7} {\bibfield
  {journal} {\bibinfo  {journal} {J. Mol. Spectrosc.}\ }\textbf {\bibinfo
  {volume} {65}},\ \bibinfo {pages} {65} (\bibinfo {year} {1977})}\BibitemShut
  {NoStop}%
\bibitem [{\citenamefont {Dantus}(2024)}]{Dantus:Science385:eadk1833}%
  \BibitemOpen
  \bibfield  {author} {\bibinfo {author} {\bibfnamefont {M.}~\bibnamefont
  {Dantus}},\ }\bibfield  {title} {\bibinfo {title} {Ultrafast studies of
  elusive chemical reactions in the gas phase},\ }\href
  {https://doi.org/10.1126/science.adk1833} {\bibfield  {journal} {\bibinfo
  {journal} {Science}\ }\textbf {\bibinfo {volume} {385}},\ \bibinfo {pages}
  {eadk1833} (\bibinfo {year} {2024})}\BibitemShut {NoStop}%
\bibitem [{\citenamefont {Zare}(1998)}]{Zare:Science279:1875}%
  \BibitemOpen
  \bibfield  {author} {\bibinfo {author} {\bibfnamefont {R.~N.}\ \bibnamefont
  {Zare}},\ }\bibfield  {title} {\bibinfo {title} {{Laser Control of Chemical
  Reactions}},\ }\href {https://doi.org/10.1126/science.279.5358.1875}
  {\bibfield  {journal} {\bibinfo  {journal} {Science}\ }\textbf {\bibinfo
  {volume} {279}},\ \bibinfo {pages} {1875} (\bibinfo {year}
  {1998})}\BibitemShut {NoStop}%
\end{thebibliography}%
\onecolumngrid
\clearpage
\listofnotes
\end{document}